\documentclass[a4paper,twoside,twocolumn,english,showpacs,PRA]{revtex4}
\usepackage{times}
\usepackage[OT1]{fontenc}
\usepackage[latin1]{inputenc}
\usepackage{amsmath}
\usepackage{graphicx}
\usepackage{amssymb}

\makeatletter

\usepackage{babel}
\makeatother
\begin{document}

\title{PGPE theory of finite temperature collective modes for a trapped Bose gas.}

\author{A. Bezett}
\author{P. B. Blakie}

\affiliation{Jack Dodd Centre for Quantum Technology,
Department of Physics, University of Otago, PO Box 56, Dunedin, New Zealand}
\date{\today}

\newcommand{\etal}{\emph{et al.}} 
\newcommand{\rC}{\text{\bf{C}}} % c-field region
\newcommand{\rI}{\text{\bf{I}}} % incoherent region
\newcommand{\PC}{\mathcal{P}_{\rC}}
\newcommand{\bef}{\hat{\psi}}
\newcommand{\bfI}{\bef_{\rI}}
\newcommand{\cf}{\psi_{\rC}} 
\newcommand{\cfp}{\psi_{\rC'}} 
\newcommand{\ecut}{\epsilon_{\rm cut}} 
\newcommand{\CF}{c-field}
\newcommand{\Nc}{N_{\rm{cond}}}
\newcommand{\thold}{t_{\rm{hold}}}
\newcommand{\nmin}{n_{\min}}

\begin{abstract}
We develop formalism based on the projected Gross Pitaevskii equation to simulate the finite temperature collective mode experiments of Jin \etal\ [PRL {\bf 78}, 764 (1997)]. We examine the $m=0$ and $m=2$ quadrupolar modes on the temperature range $0.51T_c-0.83T_c$ and calculate the frequencies of, and phase between, the condensate and noncondensate modes, and the condensate mode damping rate.
This study is the first quantitative comparison of the projected Gross-Pitaevskii equation to experimental results in a dynamical regime.
\end{abstract}

\maketitle

\section{Introduction}
The response of a manybody system to an external perturbation, particularly its collective mode response, forms an important method of analysis in condensed matter physics. 
 Since the experimental realisation of dilute gas Bose-Einstein condensation (BEC) there have been several collective mode experiments  \cite{jin96,jin97,Mewes1996,ketterle,foot}, in which perturbations of the confining potential were used to excite the system. 
 Of particular interest is the 1997 experiment of Jin \etal\ at JILA \cite{jin97}, which determined the excitation frequencies for the lowest energy quadrupolar collective modes over a temperature range spanning the condensation transition.
 At low temperatures, where the system was mainly condensate, the results were accurately described by simple meanfield theory \cite{TheoryA1,TheoryA6,TheoryA10}. However, the behavior of the collective mode frequencies at higher temperature, where a significant thermal fraction was present, proved much more difficult to describe. Indeed, the description of these experiments has become the \emph{de facto} standard for testing finite temperature quantum field theories of BEC, and has been largely responsible for the development of gapless \cite{Proukakis1998a} and second order \cite{Morgan2000a} theories of the trapped Bose gas.
 To date, the only fully quantitative theoretical descriptions of these results have been provided by the Zaremba-Nikuni-Griffin (ZNG) formalism calculations of Jackson \etal\ \cite{TheoryB3} in 2002 and the second order theory of Morgan \etal\ \cite{TheoryA5} in 2003.

\begin{figure}
\includegraphics[width=3.3in, keepaspectratio]{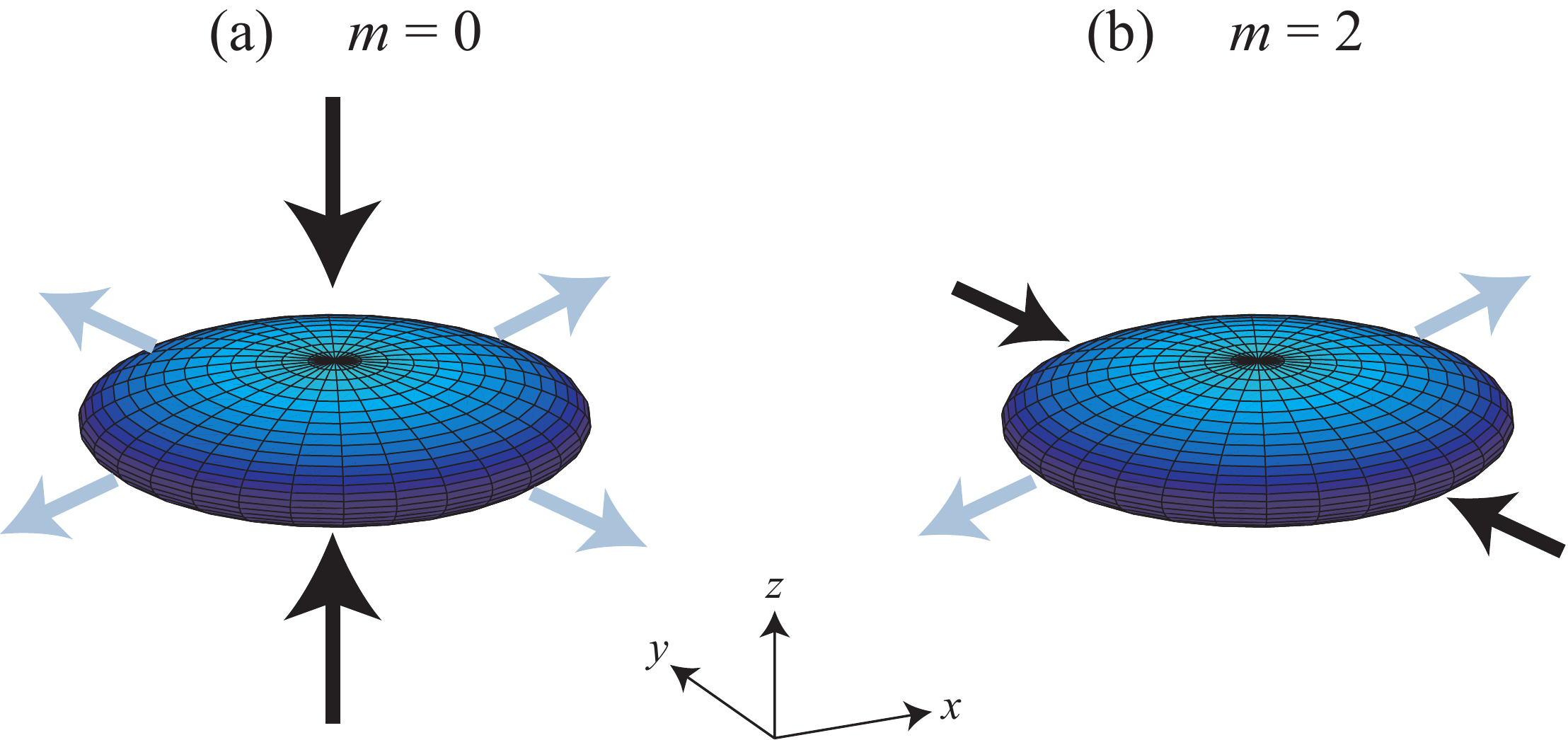}
\caption{\label{Qmodes}  Schematic view of the quadrupolar modes of an oblate condensate. (a) The $m=0$ excitation where the axial and radial widths of the condensate oscillate out of phase and (b) the $m=2$ excitations and the widths of the condensate in the $x$ and $y$ directions oscillate out of phase.}
\end{figure}

In Fig. \ref{Qmodes}(a) and (b) we show a schematic representation of the quadupolar modes excited in the JILA experiment, characterized by the projection, $m$, of their angular momentum onto the $z$ axis.
A large body of theoretical work has been conducted on the subject of the JILA experiments \cite{Proukakis1998a,TheoryA2,TheoryA3,TheoryA4,TheoryA5,TheoryA7,TheoryA8,Morgan_full,TheoryB1,TheoryB2,TheoryB3,TheoryB4,Morgan2000a,Liu2004a,Geddes2005a} and is nicely summarized in a recent review by Proukakis and Jackson \cite{Proukakis2008a}. The temperature dependence of the $m=2$ mode was accounted for by gapless Hartree-Fock Bogoliubov (HFB) theory calculations in 1998 \cite{TheoryA2}, which included anomalous average and manybody effects in the system description (also see Refs. \cite{Minguzzi1997,Shi1999,Reidl2000,Giorgini2000}). However, gapless HFB failed to account for the rather sudden upward shift in the $m=0$ mode frequency observed in experiments at $T\approx0.65T_c$. An explanation for the unexplained behavior of the $m=0$ mode was first provided by Stoof and coworkers \cite{TheoryB1,TheoryB5} (also see \cite{Storey1998}), who suggested that it arose from the coupling of in-phase and out-of-phase  oscillations of the condensate and thermal cloud. This hypothesis suggested that an adequate theoretical description would require a dynamic treatment of both the condensate and noncondensate parts of the system. 
The first such formalism was the ZNG finite temperature theory  \cite{ZNG,ZNG1,ZNG2} in which the system  description takes the form of a Gross-Pitaevksii equation for the condensate, coupled 
to a Boltzmann equation for the noncondensate.  Jackson and Zaremba \cite{TheoryB3} applied the ZNG theory to model the JILA experiment   and found relatively good agreement with the experimental results. 
The following year, Morgan \etal\ \cite{TheoryA5} reported  the results of a second order theory that were also in good agreement with the experimental results. That theory, the culmination of seven years of work by Burnett, Hutchinson, Morgan,  Proukakis and coworkers \cite{ Proukakis1998a,TheoryA2,TheoryA3,TheoryA4,TheoryA5,TheoryA7,TheoryA8,Morgan2000a,Morgan2004a,Morgan_full}, consistently included the dynamical interactions between the condensate and noncondensate atoms.

In this paper we develop the projected Gross-Pitaevskii equation (PGPE) formalism to model the experiment of Jin \etal\ \cite{jin97}. The PGPE method is a \emph{c-field} technique \cite{cfieldRev2008} applicable to the study of finite temperature degenerate Bose gases. It includes interactions between low energy modes of the gas non-perturbatively  and is applicable in the critical region, e.g. see \cite{DavisTemp,bezett,Simula2006a,Simula2008a}. Indeed, PGPE predictions for the shifts in critical temperature \cite{Davis06} are in good agreement with experimental measurements \cite{Gerbier04}. While this formalism has successfully predicted equilibrium properties for a degenerate Bose cloud, there have been no quantitative comparisons to dynamical experiments, so our comparison to the experiments of Jin \etal\ \cite{jin97} forms an important test of this theory.

\begin{figure}
\includegraphics[width=3.3in, keepaspectratio]{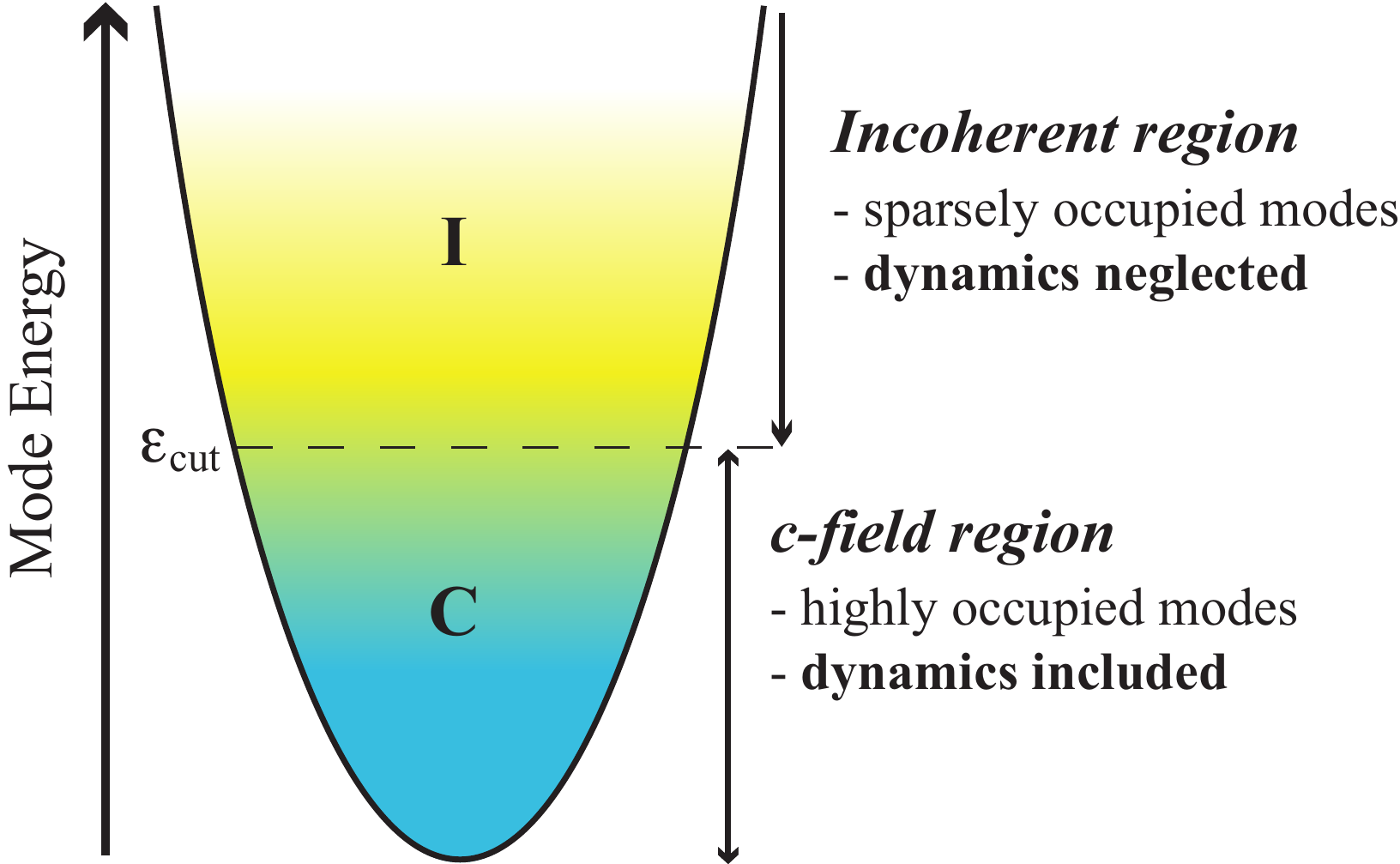}
\caption{\label{regions}  Schematic view of the \CF\ and the incoherent regions for a Bose gas in a harmonic trap potential, and the approximations we employ in our treatment of the collective mode dynamics.}
\end{figure}

A central feature of the PGPE approach is the formal division of the system modes into the classical region $\rC$ \cite{Blakie2007CR}, which is simulated using the PGPE, and an incoherent region $\rI$, for which a classical field treatment is inappropriate (see Fig. \ref{regions}). The $\rC$ region dynamics are accounted for in the PGPE description, and the $\rI$ region dynamics could be treated using, e.g., a Boltzmann description.
As previous theoretical work has has shown, the full dynamical treatment of the noncondensate is crucially important in providing a correct description of the JILA experiments. However, a full dynamical treatment of the $\rI$ region is a rather complex addition to the theory that we do not consider here. Instead,  we simply ignore the dynamics of the $\rI$ region, with the justification that many of the noncondensate modes exist in the $\rC$ region and their dynamical effect is included in the PGPE.
We critically analyze this approximation by quantifying the dependence of equilibrium and dynamic properties on the energy cutoff, $\ecut$, which sets the division between the $\rC$ and $\rI$ regions.

The organisation of the paper is as follows. In Sec. \ref{Formalism} we review the PGPE formalism for equilibrium properties of a trapped Bose gas, before outlining our extensions to the theory to model the JILA collective mode experiments. The results of our calculations for the equilibrium states, collective mode frequencies, and damping rates are presented in Sec. \ref{Results}. In that section we also consider the relative phase between the condensate and noncondensate modes, and the cutoff sensitivity of our predictions, before we conclude in Sec. \ref{Conclusion}. The data used to prepare our initial states is summarized in Appendix \ref{Aparams}.

\section{Formalism}\label{Formalism}
We take our system to be described by  the second quantized Hamiltonian
\begin{eqnarray}
\hat{H} &=& \int  d\mathbf{r}\, \left\{\hat\Psi^{\dagger}(\mathbf{r}) \left(-\frac{\hbar^2}{2m}\nabla^2 + V(\mathbf{r},t)\right)\hat\Psi(\mathbf{r})\right. \nonumber\\
&& \left.+ \frac{1}{2}U_0\hat\Psi^{\dagger}(\mathbf{r})\hat\Psi^{\dagger}(\mathbf{r})\hat\Psi(\mathbf{r})\hat\Psi(\mathbf{r})\right\} ,
\end{eqnarray}
where $\hat\Psi(\mathbf{r})$ is the quantum Bose field operator, and $U_0 = 4\pi\hbar^2a/m$ is the interaction strength, with $a$ the s-wave scattering length. The trap potential is given as
\begin{equation}
V(\mathbf{r},t) = V_0(\mathbf{r)} + \delta V (\mathbf{r},t),
\end{equation}
where 
\begin{equation}
V_0(\mathbf{r}) = \frac{1}{2}m\left(\omega_x^2x^2+\omega_y^2y^2+\omega_z^2z^2\right),
\end{equation}
is the static harmonic trapping potential, and $\delta V (\mathbf{r},t)$ is a time-dependent perturbing potential we discuss further below.

\subsection{Experimental procedure}
The theory we develop here is relevant to the finite temperature excitation experiment undertaken by Jin \etal\ in Ref. \cite{jin97}. In that experiment a  degenerate $^{87}$Rb Bose gas was prepared in a magnetic trap with frequencies $\omega_r\equiv\omega_{x,y}=2\pi\times129$Hz, $\omega_z=2\pi\times365$Hz, and initial temperatures ranging from $0.4$--$1.4 T_c$. The total number of atoms increased with temperature, varying from about $5\times10^3$ to $60\times10^3$ atoms over that temperature range, with a condensate number of about $6\times10^3\pm2\times10^3$ for $T\lesssim0.9T_c$.

Two different symmetries of perturbation were investigated in experiments, chosen to effectively couple to the lowest energy $m=0$ and $m=2$ collective modes. 
To excite the collective mode the trap was perturbed for $14$ ms and then evolved in the static trap for a variable hold time before the cloud was released and imaged after expansion (see Fig. \ref{pert}(a)). The condensate and non-condensate components were determined using bimodal fits to the absorption image, and the widths of each component were extracted as a function of time. These results were analyzed to give excitation frequencies and damping rates for both components.

\subsubsection{Time dependent perturbation}
\begin{figure}
\includegraphics[width=3.3in, keepaspectratio]{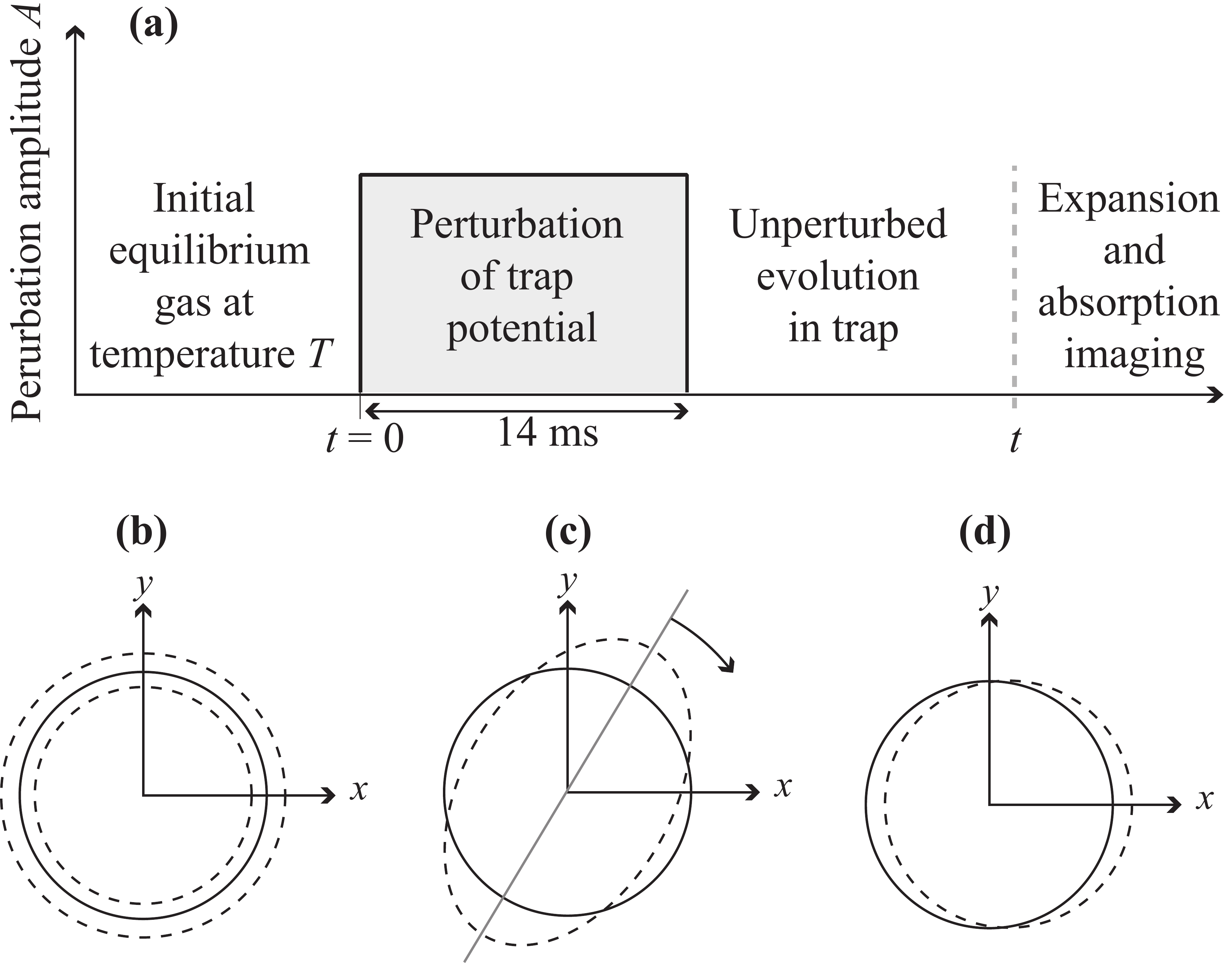}
\caption{\label{pert} 
Experimental time sequence and form of trap perturbations used to excite the Bose gas.
(a) Overview of time sequence used in the experiment to excite and observe collective excitations in the system. (b)-(d) Schematic representations of the various trap perturbations used (see text).  Contours of equipotential in the $xy$-plane are shown for the unperturbed (solid lines) and for the perturbed (dashed lines) traps. (b) Symmetric perturbation used to drive the $m=0$ mode. (c) The perturbation used to drive the $m=2$ mode corresponds to a rotating ellipse. (d) Trap center displacement used to drive the dipole mode. }
\end{figure}

The perturbation used to drive the $m=0$ mode was a weak sinusoidal modulation of the radial trap frequency (see Fig. \ref{pert}(b)).
For the $m=2$ mode the trap frequencies in the $x$ and $y$ directions were modulated sinusoidally with $\pi$ phase difference (see Fig. \ref{pert}(c)).
For calibration, the dipole mode was also measured by centre-of-mass excitation (see Fig. \ref{pert}(d)).

In the our approach to modeling these collective excitations we explicitly simulate the perturbation procedure used in experiments. To do this we use a perturbation potential of the form 
\begin{equation}
\delta V (\mathbf{r},t) = \frac{m}{2}A(t)\left\{\omega_x^2x^2\cos(\omega_p t + \phi) + \omega_y^2y^2\cos(\omega_p t) \right\},\label{pert1}
\end{equation}
where $\omega_p$ is the perturbation frequency, $\phi$ is a phase factor between the $x$ and $y$ perturbation, and $A(t)$ is the dimensionless time dependent amplitude of the perturbation (see  Fig. \ref{pert}(a)) of the square pulse form
\begin{equation}
A(t)=\left\{
\begin{array}{cc}
 A_0,  & 0\le t\le 14 \,\rm{ms} ,   \\
 0, &  \rm{otherwise},
\end{array}
\right.\label{perttiming}
\end{equation}
with $A_0=0.015$.
  The choice of $\phi=0$ ($\phi=\pi$) in Eq. (\ref{pert1}) corresponds to the perturbation used in experiment to excite the $m=0$ ($m=2$) mode.
In experiment $\omega_p$ was chosen ``to match the frequency of the excitation being studied", with the motivation that this should cause the system to oscillate at its natural frequency.

%The amplitude of perturbation $A(t)$ is varied as is done in \cite{jin97}, with the trap perturbation turned on for $15ms$ at a fixed strength, and then turned off for the observation period of $18ms$. We probe the cloud evolution in situ. The quadrupole mode excited depends on the perturbation frequency $\omega_p$ and the phase factor $\phi$ that are chosen.  When $\phi$ is zero, the corresponding mode is a quadrupole mode that breathes in the $xy$ plane, with out of phase motion in the $z$ direction. When $\phi$ is $\pi$, the corresponding mode has an ellipse that rotates in the $xy$ plane. To mimic the experiment \cite{jin97} as closely as possible, we choose $\omega_p$ to be the center frequency of the excitation to be studied. Perturbing the trap for a short time with frequency close to that measured should cause the system to oscillate at its natural frequency, as was noted in \cite{jin97}.  It is also of interest to study the dipole mode of the system. To excite the dipole mode in the $x$ direction, we introduce the following perturbation

To drive the dipole oscillation, we use a perturbation potential of the form
\begin{equation}
\delta V (\mathbf{r},t) = \frac{1}{2}m\omega^2_x \{d^2\sin^2(\omega_x t) - 2xd\sin(\omega_x t)\},\label{pert2}
\end{equation}
where $d=0.034\mu$m is the amplitude of the sinusoidal motion of the trap in the $x$ direction.

\subsection{PGPE formalism} \label{secPGPEformalism}
We briefly outline the projected Gross-Pitaevskii equation (PGPE) formalism, which is developed in detail in Ref. \cite{PGPE}.  
The Bose field operator is split into two parts according to
\begin{equation}
\hat\Psi(\mathbf{r}) = \cf(\mathbf{r}) + \bfI(\mathbf{r}),\label{EqfieldOp}
\end{equation}
where $\cf$ is the coherent region \CF\  and $\bfI$ is the incoherent field operator (see \cite{cfieldRev2008}).
These fields are defined as the low and high energy projections of the full quantum field operator, separated by the energy $\ecut$, as shown in Fig. \ref{regions}. In our theory this cutoff is implemented in terms of the harmonic oscillator eigenstates $\{\varphi_n(\mathbf{r})\}$ of the time-independent single particle Hamiltonian
\begin{equation}
H_0=-\frac{\hbar^2}{2m}\nabla^2+V_0(\mathbf{r}),
\end{equation}
i.e. $\epsilon_n\varphi_n(\mathbf{r})=H_0\varphi_n(\mathbf{r})$, with $\epsilon_n$ the respective eigenvalue.
The fields are thus defined by
\begin{eqnarray}
\cf(\mathbf{r}) &\equiv&\sum_{n\in\rC}c_n\varphi_n(\mathbf{r}),\\
\bfI(\mathbf{r}) &\equiv&\sum_{n\in\rI}\hat{a}_n\varphi_n(\mathbf{r}),
\end{eqnarray}
where the $\hat{a}_n$ are Bose annihilation operators, the $c_n$ are complex amplitudes, and the sets of quantum numbers defining the regions are 
\begin{eqnarray}
\rC &=&\{n:\epsilon_n\le \ecut\},\\
\rI &=&\{n:\epsilon_n> \ecut\}. 
\end{eqnarray} 
The applicability of the PGPE approach to describing the finite temperature gas relies on an appropriate choice for $\ecut$, so that the modes at the cutoff have an average occupation of order unity. This choice means that the all the modes in $\rC$ are {appreciably occupied}, justifying the classical field replacement $\hat{a}_n\to c_n$. In contrast the $\rI$ region contains many sparsely occupied modes that are particle-like and would be poorly described using a classical field approximation. Here we treat these modes using a meanfield approach.

\subsection{Equilibrium states}\label{secEqstates}
In this subsection we review our procedure for calculating finite temperature equilibrium properties of a trapped Bose gas. The basic approach is to treat the $\rC$ and $\rI$ regions as independent systems in thermal and diffusive equilibrium. We discuss the treatment of these regions separately below. Further details on this procedure are given in Sec. 3 of \cite{cfieldRev2008}.

\subsubsection{PGPE treatment of $\rC$ region}

%There has been a significant body of work on applications of the classical field methods to  zero and finite temperature properties of Bose gases \cite{Davis2002a,DavisTemp,Davis2005a,Davis06,Simula2006a,Gardiner2002a,Gardiner2003a,Goral2001a,Davis2001b,Davis2001a,Lobo2004a,Marshall1999a,Norrie2004a,
%Polkovnikov2004a,Sinatra2001a,Sinatra2002a,Steel1998a,Bradley,Blakie2004}. These studies  show that this splitting of the field operator can be done in a consistent manner and provides an accurate description of experiment (in particular see Ref. \cite{Davis06}). We apply the classical field approximation to the coherent field operator, thus neglecting quantum fluctuations. The coherent field can be expanded in a harmonic oscillator basis
%\begin{equation}
%\cf(\mathbf{r},t)=\sum_{n\in  C}c_n(t)\varphi_n(\mathbf{r}),
%\end{equation}
%where $n$ represents the quantum numbers for the harmonic oscillator basis states $\{\varphi_n(\mathbf{r})\}$, and $c_n$ are time-dependent complex amplitudes. The coherent region ($\rC$) is defined by restricting the sum to single particle states of energy less than $\ecut$.

The equation of motion for $\cf$ is the PGPE
\begin{eqnarray}
i\hbar\frac{\partial \cf }{\partial t} = H_0\cf + \PC\left\{ U_0 |\cf|^2\cf\right\}, \label{PGPE}
\end{eqnarray}
where the projection operator 
\begin{equation}
\PC\{ F(\mathbf{r})\}\equiv\sum_{n\in\rC}\varphi_{n}(\mathbf{r})\int
d\mathbf{r}\,\varphi_{n}^{*}(\mathbf{r}') F(\mathbf{r}'),\label{eq:projectorC}\\
\end{equation}
formalises our basis set restriction of $\cf$ to the $\rC$ region. The main approximation used to arrive at the PGPE is to neglect dynamical couplings to the incoherent region \cite{Davis2001b}. 
%While it is expected to be well justified in the equilibrium case, the validity of this approximation in dynamical cases is uncertain. An evaluation of the applicability of this approximation to dynamic processes is a key focus of this paper.

An important feature of Eq.~(\ref{PGPE}) is that it is ergodic, so that  the microstates $\cf$ evolves through in time form a sample of the equilibrium microstates, and  time-averaging can be used to obtain macroscopic equilibrium properties.
Our basic procedure for finding equilibrium states consists of evolving the PGPE with three adjustable parameters: (i) the  cutoff energy, $\ecut$, that defines the division between $\rC$ and $\rI$, and hence the number of modes in the $\rC$ region; (ii) the number of $\rC$ region atoms, $N_{\rC}$; (iii) the total energy of the $\rC$ region, $E_\rC$. The last two quantities, defined as 
\begin{eqnarray}
E_{\rC}&=&\int d\mathbf{r}\,\cf^*\left(H_0 +  \frac{U_0}{2} |\cf|^2\right)\cf,\\
N_{\rC} &=& \int d\mathbf{r}\,|\cf(\mathbf{r})|^2,
\end{eqnarray}
are important because they represent constants of motion of the PGPE (\ref{PGPE}), and thus control the equilibrium state of the system. 

Given choices of these parameters, randomized initial states are constructed satisfying those constraints (see Sec. 3 of Ref. \cite{cfieldRev2008}), and are evolved according to Eq.~(\ref{PGPE}). Typically evolution times of order 10 trap periods are used for the system to relax towards equilibrium \cite{Blakie2008a}. Further evolution for times of order 10-100 trap periods are used to sample equilibrium microstates or to time-average equilibrium properties.

To characterize the equilibrium state in the $\rC$ region it is necessary to determine the average density, condensate fraction, temperature and chemical potential. Many of these quantities are also important for characterizing the $\rI$ region (see Sec. \ref{secMFrI}).  

The average density is obtained as a time average over the field microstates, i.e.,
\begin{eqnarray}
n_{\rC}(\mathbf{r}) &\equiv&\left\langle \left|\cf(\mathbf{r})\right|^2\right\rangle,\label{nc1}\\ &\approx& \frac{1}{M_s}\sum_{j=1}^{M_s}\left|\cf(\mathbf{r},\tau_j)\right|^2,\label{nc2}
\end{eqnarray}
where $\{\tau_j\}$ is a set of $M_s$ times (after the system has been allowed to relax to equilibrium) at which the field is sampled. We typically use $\sim500$ samples over
100 trap periods to perform such averages.
The notation $n_\rC(\mathbf{r})$ emphasizes this is the density contribution from atoms residing in the $\rC$ region, with the obvious property that  $N_\rC=\int d\mathbf{r}\,n_\rC(\mathbf{r})$.
For later convenience, we note that the momentum properties of the $\rC$ region are easily evaluated using the momentum field $\phi_{\rC}(\mathbf{p})$, obtained by Fourier transforming the spatial field, i.e.
\begin{equation}
\phi_\rC(\mathbf{p},t) = \frac{1}{\hbar^3} \int d \mathbf{r}\, e^{-i\mathbf{p}\cdot\mathbf{r}/\hbar}\cf(\mathbf{r},t).
\end{equation} 

To find the condensate number, $\Nc$, in our equilibrium state, we use the Penrose Onsager definition \cite{Penrose1956}, that  $\Nc$ is given by the largest eigenvalue of the one-body density matrix 
\begin{equation}
G^{(1)}(\mathbf{r},\mathbf{r'}) = \langle \cf^*(\mathbf{r})\cf(\mathbf{r}') \rangle,\label{G1}
\end{equation}
which we are also able to evaluate as a time-average.

Finally, using the Rugh method \cite{Rugh1997a}, we are able to determine the temperature ($T$) and chemical potential ($\mu$) by time averaging. We refer to Refs. \cite{DavisTemp,Davis2005a} for additional details of this procedure.

\subsubsection{Meanfield treatment of $\rI$ region}\label{secMFrI}
The average properties of the incoherent region can be calculated from
the one-particle Wigner distribution\begin{equation}
W_{\rI}(\mathbf{r},\mathbf{p})=\frac{1}{\hbar^3} \frac{1}{\exp(\beta[\epsilon_{\rm{HF}}(\mathbf{r},\mathbf{p})-\mu])-1},\label{eq:Wig3D}\end{equation} 
where 
\begin{eqnarray}
\epsilon_{\rm{HF}}(\mathbf{r},\mathbf{p}) & = & \frac{p^2}{2m}+ V_0(\mathbf{r})+2U_0(n_{\rC}(\mathbf{r})+n_{\rI}(\mathbf{r})),\label{eq:EHF3D}
\end{eqnarray}
is the Hartree-Fock energy, and $\mu$ is the chemical potential.  
In this semiclassical description $\mathbf{r}$ and $\mathbf{p}$ are treated as continuous variables. However, care needs to be taken to ensure that Eq.~(\ref{eq:Wig3D}) is only applied to the appropriate region of phase space spanned by the incoherent region,  i.e.~single-particle modes of energy exceeding $\ecut$. In phase space this region is  
\begin{equation}\label{WIdef}
\Omega_{\rI}=\left\{ \mathbf{r},\mathbf{p}:\frac{p^2}{2m}+V_0(\mathbf{r})\ge\ecut\right\} .
\end{equation}
This allows us to calculate  the incoherent region density
\begin{eqnarray}
n_{\rI}(\mathbf{r})&=& \int_{\Omega_\rI}{d\mathbf{p}}\, W_{\rI}(\mathbf{r},\mathbf{p}),
\end{eqnarray}
and atom number
\begin{eqnarray}
N_\rI  &=& \int d\mathbf{r}\,n_\rI(\mathbf{r}).
\end{eqnarray}
\subsubsection{Obtaining equilibrium states}\label{sec:makeEq}
The overall algorithm for generating  equilibrium states within the PGPE formalism is summarized as a three step process:
\begin{enumerate}
\item Using selected values of $\{E_\rC$, $N_\rC,\ecut\}$, an  appropriate randomized state is constructed and evolved according to the PGPE. Using time averaging $\Nc, n_\rC(\mathbf{r}), T,$ and $\mu$ are calculated.
\item Using $n_\rC(\mathbf{r}), T,$ and $\mu$, the incoherent region is analyzed, yielding $n_{\rI}(\mathbf{r})$, and hence the total atom number 
\begin{equation}N=N_\rC+N_\rI.\end{equation}
\item The values obtained for $N$, $T$, and $\Nc$ are compared to the desired values and the  values of $\{E_\rC$, $N_\rC,\ecut\}$ are adjusted before returning to step 1.
\end{enumerate}
This process is quite time consuming since steps 1 and 2 can take of order a day to complete on commodity PC hardware. Finding desired initial configurations is made somewhat easier in the stochastic projected Gross-Pitaevskii equation (SPGPE) formalism \cite{Gardiner2002a,Gardiner2003a,cfieldRev2008} which allows direct control of $T$ and $\mu$ rather than $E_{\rC}$ and $N_{\rC}$, although we have not used that here.

Following this procedure, we have sampled equilibrium configurations with  condensate occupation $\Nc \approx 6000 \pm 2000$ over the temperature range $0.51T_c$ to $0.83T_c$ (see Sec. \ref{sec:initstates_results}).   These samples of the equilibrium state are used as initial conditions for the collective mode excitation procedure we discuss next.
We examine attributes of the initial states, particularly the dependence on $\ecut$, later in this paper.

\subsection{Formalism for dynamical modeling of collective mode excitation}
The fundamental approximation in our treatment of collective mode excitation is to neglect the the dynamics of the $\rI$ region, and their influence on the $\rC$ region.
In this approximation the many-body dynamics is described by the PGPE
\begin{eqnarray}
i\hbar\frac{\partial \cf }{\partial t} &=& H_0\cf+ \PC\left\{ \left(\delta V(\mathbf{{r}},t) +U_0 |\cf|^2\right)\cf\right\}, \label{PGPE2}
\end{eqnarray}
which differs from the PGPE used to generate equilibrium states by the inclusion of the perturbation potential.
The motivation for considering only the $\rC$ region is that many noncondensate modes and their effect on the collective mode dynamics are included in $\rC$. We critically examine this approximation  later.

\subsubsection{Initial conditions}
We begin our simulations at $t=0$ when the perturbation potential is first applied (see Eqs. (\ref{pert1}) and (\ref{perttiming})). 
The perturbation potential, and the ensuring dynamics it generates, break the ergodicity of the PGPE for some period of  time after the perturbation has concluded (until the system   rethermalized when the collective modes have damped out).
Ensemble averages of the dynamical system thus need to be taken as an average over many trajectories. For each trajectory, we take as an initial condition
\begin{equation}
\cf^{(j)}(\mathbf{r},t=0)=\cf^{\rm{Eq}}(\mathbf{r},\tau_j),
\end{equation}
where we have used the notation $\cf^{\rm{Eq}}$ to represent the equilibrium states generated for the time independent potential (i.e. the states $\cf(\mathbf{r},\tau_j)$ appearing in Eq. (\ref{nc2})), and $\cf^{(j)}$ to represent the $j$-th trajectory for the simulation of the collective mode dynamics. The subsequent evolution of $\cf^{(j)}(\mathbf{r},t)$ is according to Eq. (\ref{PGPE2}), which excites collective modes in the system .
 
\subsubsection{Observations and analysis}
 In the analysis of the system dynamics we present in the next section we make extensive use of the line density, defined for the $j$-th trajectory as
 \begin{equation}
n_l^{(j)}(x,t) = \int dy dz\, |\cf^{(j)}(\mathbf{r},t)|^2.\label{posld}
\end{equation}
This quantity, for a single trajectory, is itself of interest as the spatial integration corresponds to a spatial averaging of the system over the many modes in the $\rC$ region, and  there is some evidence that single trajectories of the PGPE can be compared to single experimental results.
However, we will also be interested in the trajectory average calculated as
 \begin{equation}
n_l(x,t) = \frac{1}{M_t}\sum_{j=1}^{M_t}n_l^{(j)}(x,t),
\end{equation}
where we use $M_t=20$   for the trajectory averaged results presented in this paper.

We will also be interested in the momentum space equivalent line densities,
\begin{eqnarray}
n_l^{(j)}(p_x,t) &=& \int dp_y dp_z\, |\phi_\rC^{(j)}(\mathbf{p},t)|^2,\label{mtmld}\\
n_l(p_x,t)&=&\frac{1}{M_t}\sum_{j=1}^{M_t}n_l^{(j)}(p_x,t).
\end{eqnarray}
We emphasize that these line densities only include contributions from atoms in the $\rC$ region. 

We can also use trajectory averaging to obtain other quantities, such as the coherent condensate component of the system. We do this by extending the one-body density matrix to the nonequilibrium case and evaluating it with trajectory averaging, i.e.
\begin{equation}
G^{(1)}(\mathbf{r},\mathbf{r'},t) = \sum_{j=1}^{M_t}\left(\cf^{(j)}(\mathbf{r},t)\right)^*\cf^{(j)}(\mathbf{r}',t).\label{G1traj}\end{equation}
Diagonalising $G^{(1)}(\mathbf{r},\mathbf{r}',t)$ at each time we can obtain the instantaneous condensate (coherent)  field $\psi_{\rm{cond}}(\mathbf{r},t)$, and hence the condensate line density
 \begin{equation}
n_l^{\rm{cond}}(x,t) = \int dy dz\, |\psi_{\rm{cond}}(\mathbf{r},t)|^2.
\end{equation}

\section{Results}\label{Results}
In this section, we present a detailed analysis of the PGPE simulations of the  JILA experiment \cite{jin97}. 
First, in Sec. \ref{sec:initstates_results}, we present the parameters of the equilibrium states we have generated that we use as the basis for our collective excitation modeling.
Then in Sec. \ref{sec:den_results},  we develop convenient observables and examine the density response of the Bose cloud to the perturbative drive. 
 In Sec. \ref{sec:freq_results} we present results for the frequencies and decay rates of the lowest energy $m=2$ and $m=0$ modes. 
Then, in Sec. \ref{sec:ecut_results}, we analyse the effect of the energy cutoff in our formalism, and provide evidence for how it affects the equilibrium and dynamic properties of the Bose cloud. For completeness, we then calculate the frequencies of the dipole mode as a function of temperature in Sec. \ref{sec:dipole_results}, and lastly discuss the phase of the noncondensate and condensate oscillations in Sec. \ref{sec:phase_results}.

\subsection{Equilibrium states}\label{sec:initstates_results}
\begin{figure}
\includegraphics[width=3.5in, keepaspectratio]{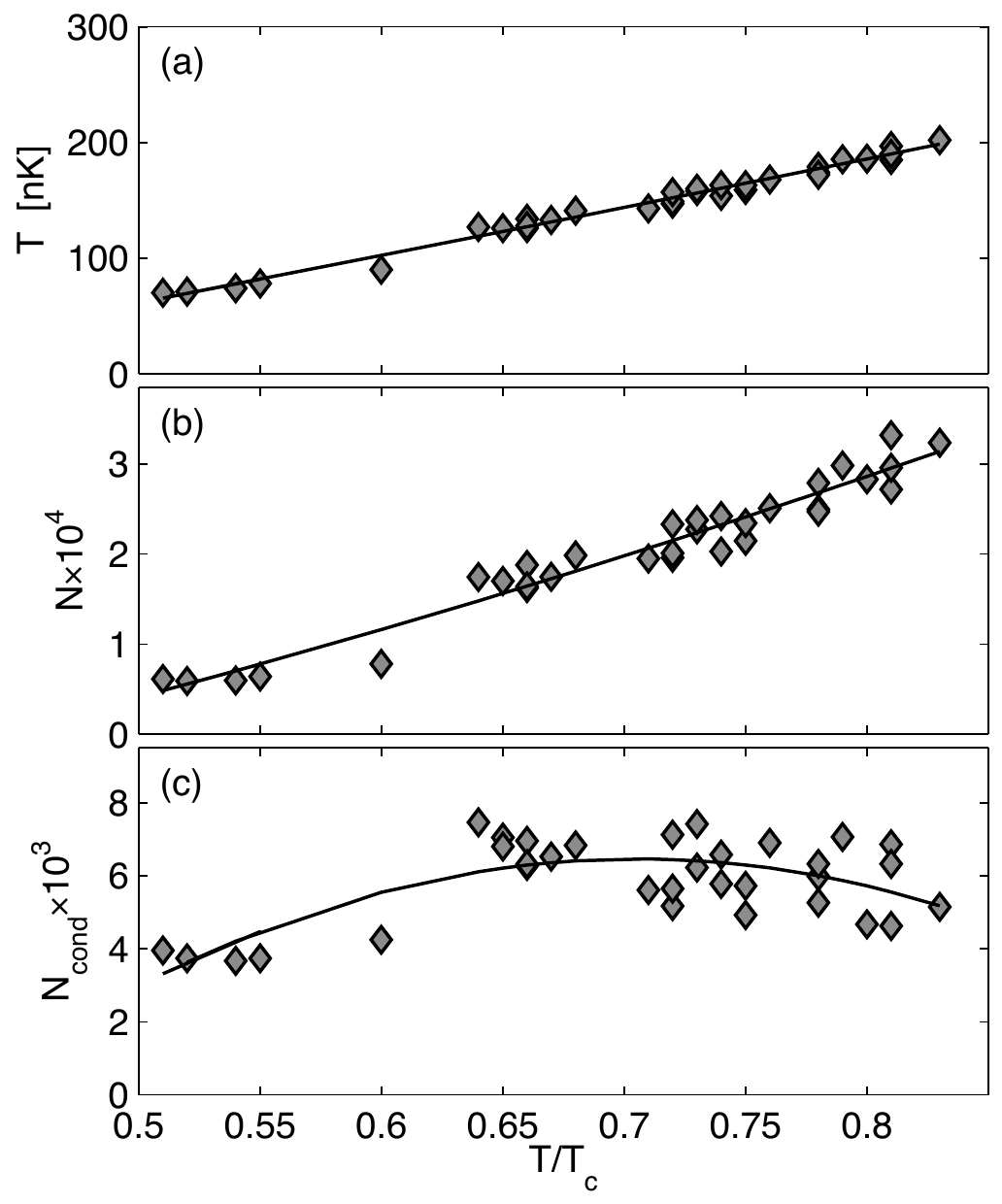}
\caption{\label{macroparams} Equilibrium state properties. (a) Temperature, (b) total atom number, and (c) condensate number as a function of $T/T_c$. PGPE results (diamonds) and lines are guides to the eye. }
\end{figure}
First we present a summary of our results for the equilibrium states generated according to the procedure discussed in Sec. \ref{sec:makeEq}. 
The macroscopic parameters of the states we have produced are shown in Fig. \ref{macroparams}. These states provide initial conditions over the temperature range $0.51T_c-0.83T_c$ with a condensate number in the range $3.5\times10^3-7.5\times10^3$, which is comparable to the spread in condensate values used in experiment over this temperature range (see. Fig. 1(c) of Ref. \cite{jin97}).
A complete list of the parameters and properties of our initial equilibrium states is given in Appendix \ref{Aparams}.

\subsection{Density response}\label{sec:den_results}
In this section, we show examples of the density response of the system after the sinusoidal perturbation has been switched off, and the cloud is evolving \emph{in situ} in a static harmonic potential.

\begin{figure}
\includegraphics[width=3.5in, keepaspectratio]{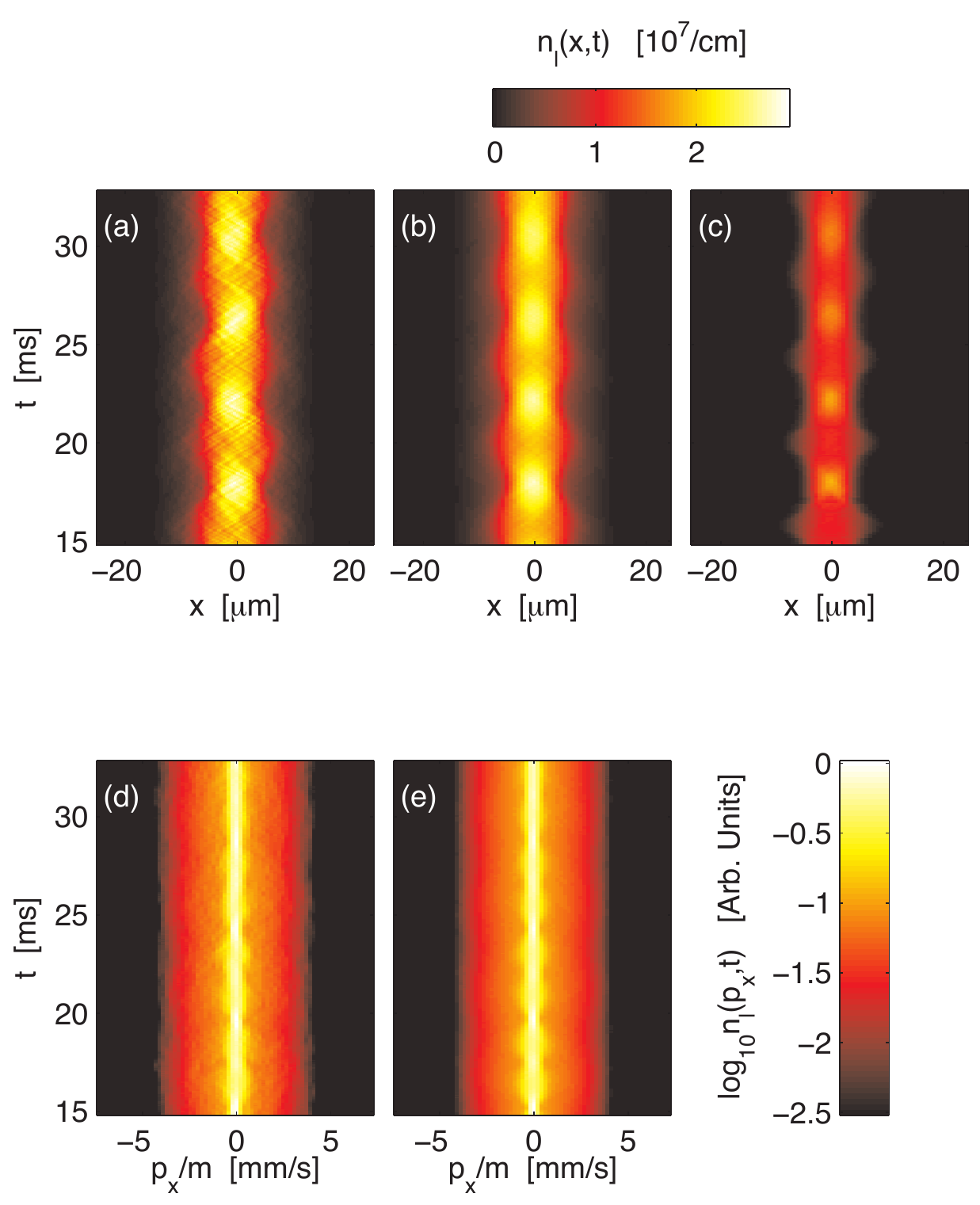}
\caption{\label{pos_mesh} Line densities evolution after perturbation. (a) A single trajectory position space line density   $n_l^{(j)}(x,t)$.  (b) Trajectory averaged position space line density   $n_l(x,t)$. (c) Condensate position space line density   $n_l^{\rm{cond}}(x,t)$.  (d) A single trajectory momentum space line density   $n_l^{(j)}(p_x,t)$.  (e) Trajectory averaged momentum space line density   $n_l(p_x,t)$. Results for a system with $T=154$ nK, $N=2.0\times10^4$, $\Nc = 5.7\times10^3$ }
\end{figure}
\begin{figure}
\includegraphics[width=3.5in, keepaspectratio]{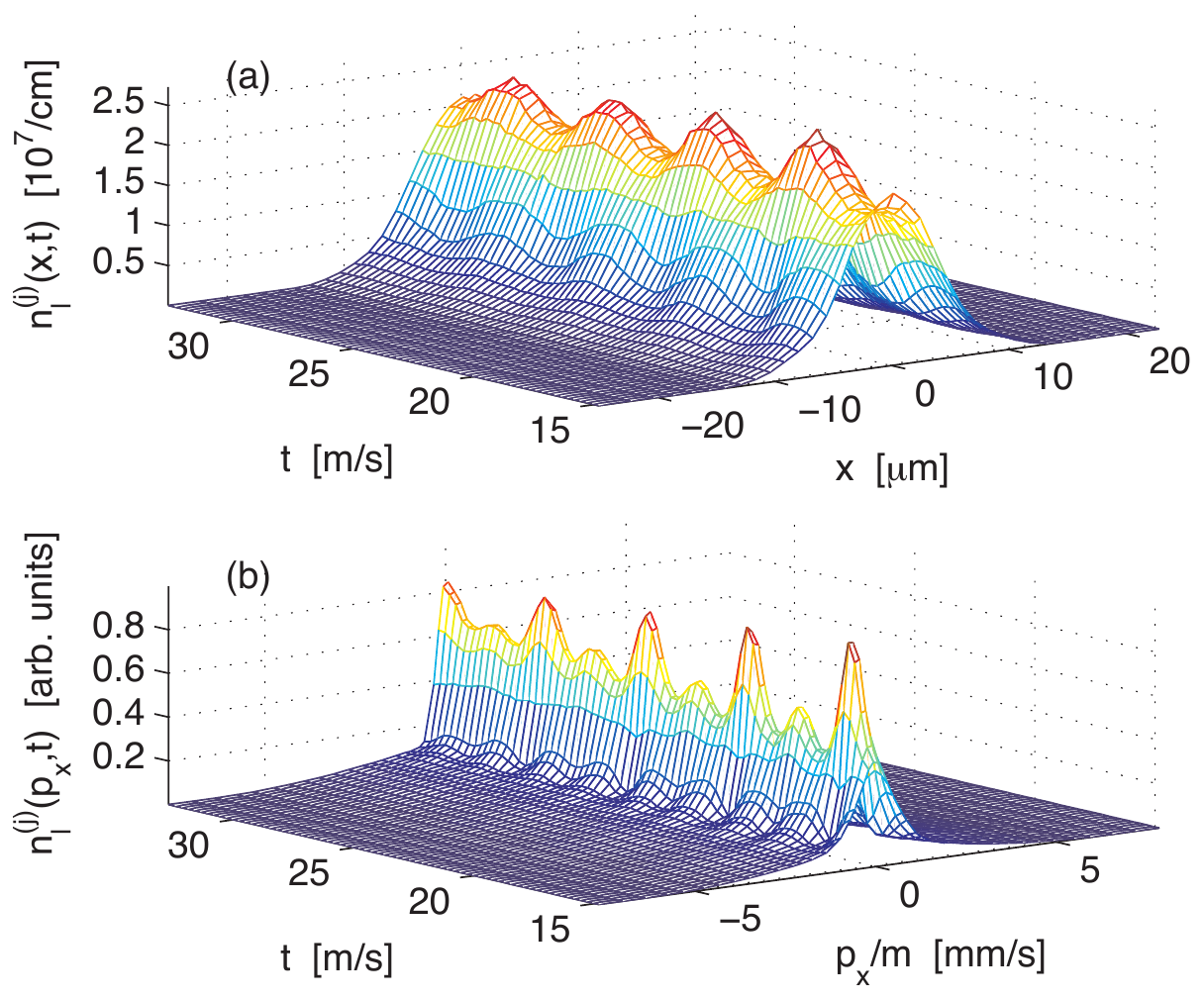}
\caption{\label{pos_mesh2} Surface plots of trajectory averaged line density plots. (a)  Trajectory averaged position space line density   $n_l(x,t)$.  (b) Trajectory averaged position space line density   $n_l(p_x,t)$. Same data as displayed in Fig. \ref{pos_mesh}(b) and (e)}
\end{figure}
\begin{figure}
\includegraphics[width=3.5in, keepaspectratio]{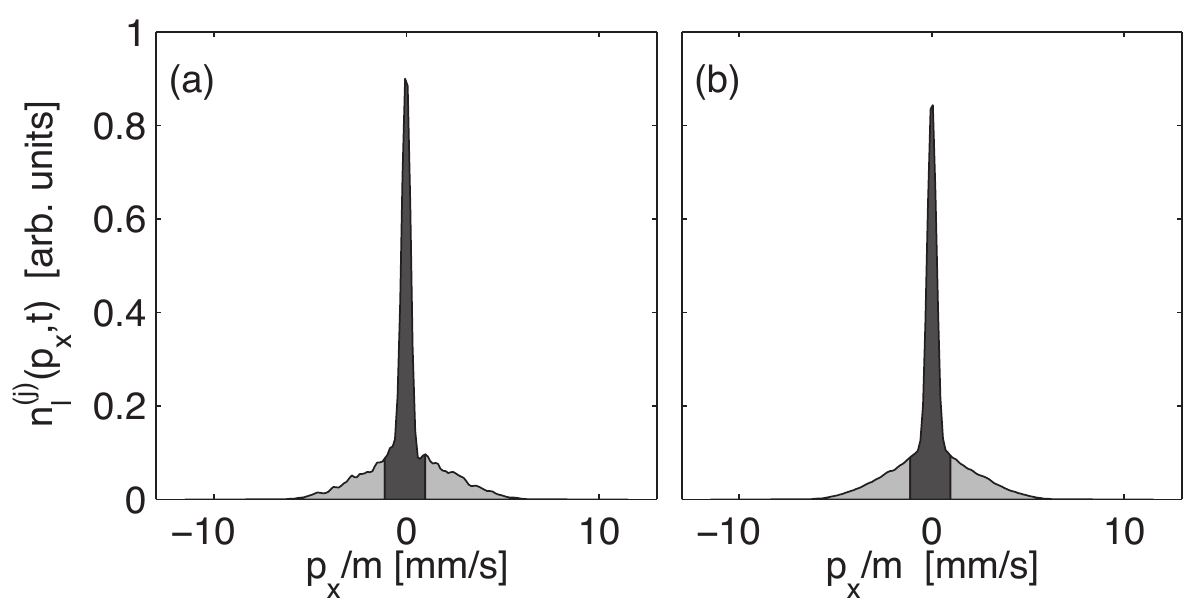}
\caption{\label{mtmslice}   (a) Single trajectory and (b) trajectory averaged momentum line density at $t=32$ ms for the condensate domain (dark shaded region) and the noncondensate domain (light shaded region) are shown (see text). Results for the same parameters given in Fig. \ref{pos_mesh}. }
\end{figure}
 
Figures \ref{pos_mesh}(a)-(e) and Figs. \ref{pos_mesh2}(a)-(b) show the evolution of the position and momentum line densities for a  Bose gas after the perturbation with $m=0$ symmetry has been applied, where the time is measured with $t=0$ corresponding to the beginning of the perturbation (see Fig. \ref{pert}). 
The timescale of these results corresponds to the period of observation used in experiments.
The position line density has a clear width oscillation induced by the perturbation. We have made similar observations of the $y$ and $z$ line densities (defined analogously to Eqs. (\ref{posld}) and (\ref{mtmld})) and have verified that width oscillations also occur. For the $m=0$ symmetry perturbation we find that the $x$ and $y$ oscillations are in phase, whereas the $x$ and $z$ oscillations are out of phase. Thus we conclude that the perturbation has excited the $m=0$ mode more strongly than any other mode.

A similar study of the density response of the system to the perturbation with $m=2$ symmetry reveals expected behavior: the widths in the $x$ and $y$ directions oscillate out of phase, and the $z$ width remains (approximately) constant.

Figure \ref{pos_mesh}(d) shows the momentum line density for a single trajectory of a Bose gas after the perturbation with $m=0$ symmetry has been applied. Figure \ref{pos_mesh}(d)  and Fig. \ref{pos_mesh2}(b) show the trajectory averaged line density.
The momentum line density is sharply peaked at $p_x=0$ due to the presence of a condensate. The peak value of the momentum line density oscillates periodically with minor peaks occurring between major peaks (see Fig. \ref{pos_mesh2}(b)).
The major peak occurs first at $t\approx16$ ms and then returns each time the condensate width reaches the outer turning point of its oscillation in position space (i.e. the condensate is at its widest, see Fig. \ref{pos_mesh}(c)). This connection between the position space width and momentum space peak value for the condensate arises through the Heisenberg relationship, i.e. the position and momentum widths of the condensate mode are inversely related.
The intermediate minor peak arises because of the out of phase oscillation of condensate width in the different directions integrated over to obtain the line density\footnote{In the case of the $m=0$ mode the out-of-phase oscillations is along the $z$ direction, whereas for the $m=2$ mode is it along the $y$ direction (e.g. see Fig. \ref{Qmodes})}.

In addition to the dominant condensate peak at $p_x=0$, a broad background feature is apparent in the momentum density at larger $|p_x|$ values. This feature, which we attribute to the non-condensate portion on the system in the $\rC$ region, is more clearly apparent in  momentum line density shown in Figs. \ref{mtmslice} (a) and (b).

\subsection{Frequencies and decay rates of collective modes} \label{sec:freq_results}
\subsubsection{Observables}\label{secobservables}
It is necessary to measure appropriate observables to determine the frequencies and damping rates of collective modes excited. In experiments bimodal fitting of the expanded system provided such observables, and gave independent information for the condensate and noncondensate (or thermal cloud). 
The \emph{in situ} momentum distribution  (e.g. see the momentum line density in Fig. \ref{mtmslice}(b)) clearly reveals the distinct character of the condensate and noncondensate components, and (like in experiments) fitting a bimodal distribution to determine the widths of the condensate and thermal components would seem to be an obvious choice for observable. However, the \emph{in situ} condensate momentum peak is extremely narrow and we have found that performing bimodal fits to the momentum line density is ambiguous and noisy.  We note that in experiments the expansion procedure gives rise to considerable broadening of the condensate momentum distribution (e.g. see \cite{Stenger1999}) and thus cannot be compared directly to our \emph{in situ} line density. 

However, we can develop two useful observables that avoid the need for fitting. First we observe that the two momentum domains, defined as
\begin{eqnarray}
\rm{c} &=& \{ p_x : |p_x| \leq p_0 \},\\
\rm{n} &=& \{ p_x : |p_x| > p_0 \},
\end{eqnarray}
with $p_0=\sqrt{2\hbar m \omega_x}$,
are dominated by the condensate (i.e. narrow peak) and noncondensate (broad background) respectively (see shaded regions in Figs. \ref{mtmslice} (a) and (b)). We thus refer to these domains as the condensate (i.e. $\rm{c}$) and noncondensate (i.e. $\rm{n}$) domains respectively. The value of $p_0$ is in some sense arbitrary as long as it is greater than the condensate momentum width, and much less that the characteristic thermal momentum ($p_{\rm{th}}=h/\lambda_{\rm{db}}$, with $\lambda_{\rm{db}}$ the thermal de Broglie wavelength). Our choice, $p_0=\sqrt{2\hbar m \omega_x}$,  satisfies both of these these criteria. 
 
We can now define our two observables, as the variance of the momentum line densities on these restricted domains, i.e.
\begin{eqnarray}
P_{\rm{c}}(t) &=& \langle p_x^2 \rangle_{\rm{c}}-  \langle p_x \rangle_{\rm{c}}^2,\label{cond_reg}\\ 
P_{\rm{n}}(t) &=& \langle p_x^2 \rangle_{\rm{n}} -  \langle p_x \rangle_{\rm{n}}^2, \label{ncond_reg}
\end{eqnarray}
where
\begin{equation}
\langle p_x^j \rangle_{\sigma} = \frac{\int_{\sigma} dp_x n_l(p_x,t) p_x^j }{\int_{\sigma} dp_x n_l(p_x,t)},\label{restrictmtms}
\end{equation}
with $\sigma=\{\rm{c},\rm{n}\}$.
We note that the denominator of Eq. (\ref{restrictmtms}) appropriately normalizes the moments, and the choice of variance for $P_{\sigma}$,  rather than the second moment $\langle p_x^2\rangle_\sigma$, is to remove the effects of any residual center of mass motion of the system. 

\begin{figure}
\includegraphics[width=3.5in, keepaspectratio]{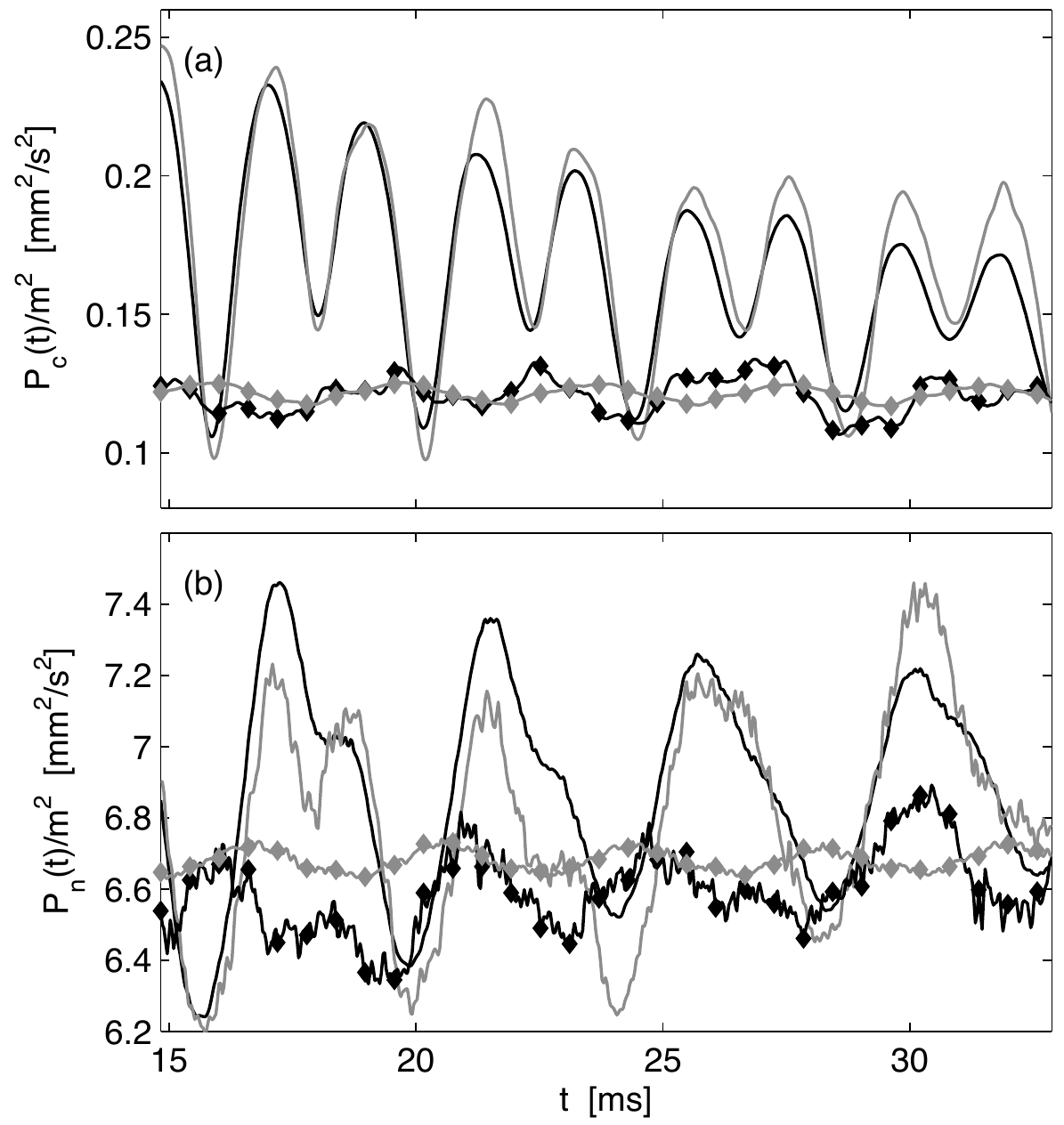}
\caption{\label{observablefig1}   (a) Condensate domain observable,  $P_{\rm{c}}(t)$, and (b) noncondensate domain observable,  $P_{\rm{n}}(t)$, as a function of time.  For a system excited by the $m=0$ symmetry perturbation: Trajectory averaged result (black line), and single trajectory result (grey line). For unperturbed equilibrium system: Trajectory averaged result (black diamond line), and single trajectory result (grey diamond line).   System parameters: $T=159$ nK$\approx0.75T_c$, $N = 2.1\times10^4$, $\Nc=4.9\times10^3$.}
\end{figure}

\begin{figure}
\includegraphics[width=3.5in, keepaspectratio]{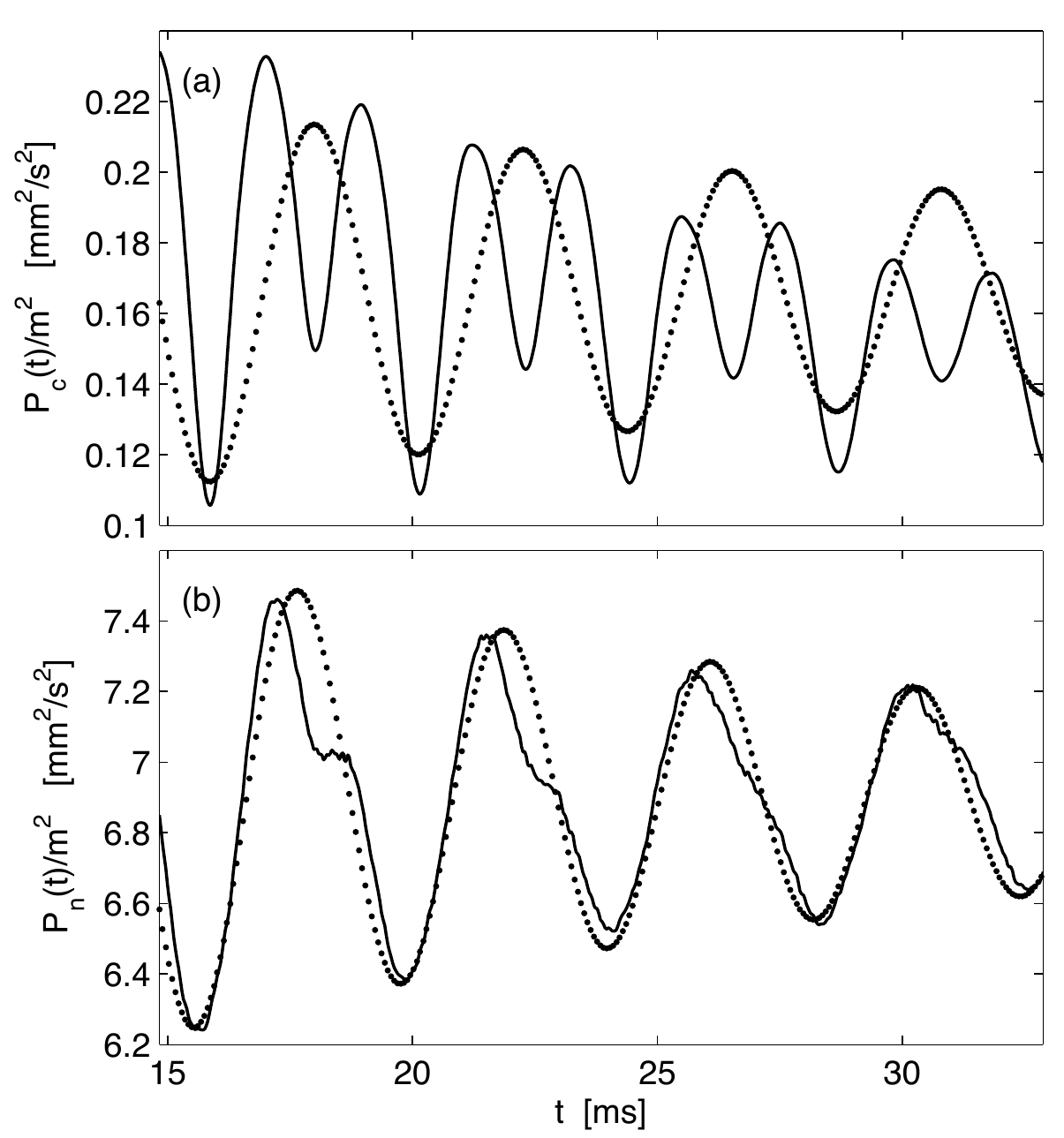}
\caption{\label{observablefigF}  Fits to observables: (a) Condensate domain observable,  $P_{\rm{c}}(t)$, (line) and (b) noncondensate domain observable, $P_{\rm{c}}(t)$,  (line) as a function of time for a system excited by the $m=0$ symmetry perturbation. Fits to observables using Eq. (\ref{Pfit}) shown (dotted lines).
 System parameters: $T=159$ nK$\approx0.75T_c$, $N = 2.1\times10^4$, $\Nc=4.9\times10^3$.}
\end{figure}

In Fig. \ref{observablefig1}(a) and (b)  we show examples of $P_{\rm{c}}(t)$ and $P_{\rm{n}}(t)$, evaluated from the PGPE simulation of an equilibrium system and a system excited by the perturbation with $m=0$ symmetry. From these results it is clear that the observables reveal the collective mode induced by the perturbation compared to the much smaller thermal fluctuations in the equilibrium states.
In the collective mode analysis we always use  $P_{\rm{c}}(t)$ and $P_{\rm{n}}(t)$ evaluated from the trajectory averaged line density, $n_l(p_x,t)$, however  the results in Fig. \ref{observablefig1} show that if the single trajectory line density, $n_l^{(j)}(p_x,t)$,  is used to evaluate these quantities a useful signal is also obtained. 

For both  $P_{\rm{c}}(t)$ and $P_{\rm{n}}(t)$ we notice that considerable damping occurs over the period of observation. In both signals anharmonic features are present, but are most apparent in the condensate observable where a weaker intermediate dip  is apparent. The origin of this feature is the same as for the intermediate peak in Fig. \ref{pos_mesh2}(b)  (see discussion in Sec. \ref{sec:den_results}): Integration over the out of phase oscillation of the $m=0$ mode in the $z$ direction. 
We have also verified that the observable signal is relatively insensitive to small adjustments of the value of $p_0$ used to define the $\rm{c}$ and $\rm{n}$ domains.

As in the experiment we fit a decaying sinusoid of the form
\begin{equation}
P_{\rm{fit}}(t)=Ae^{-\gamma t}\sin(\omega t+\vartheta)+B,\label{Pfit}
\end{equation}
 to our results, to obtain the collective mode frequency ($\omega$) and damping rate ($\gamma$). 
 Example fits to the observable, shown in Figs. \ref{observablefigF}(a) and (b), indicate that while our combination of observable and fitting function is adequate for accurately determining the mode frequency, it does not provide a good description of the amplitude or damping behavior of the modes.

\subsubsection{Mode frequencies}\label{secmodefreqs}
\begin{figure}
\includegraphics[width=3.5in, keepaspectratio]{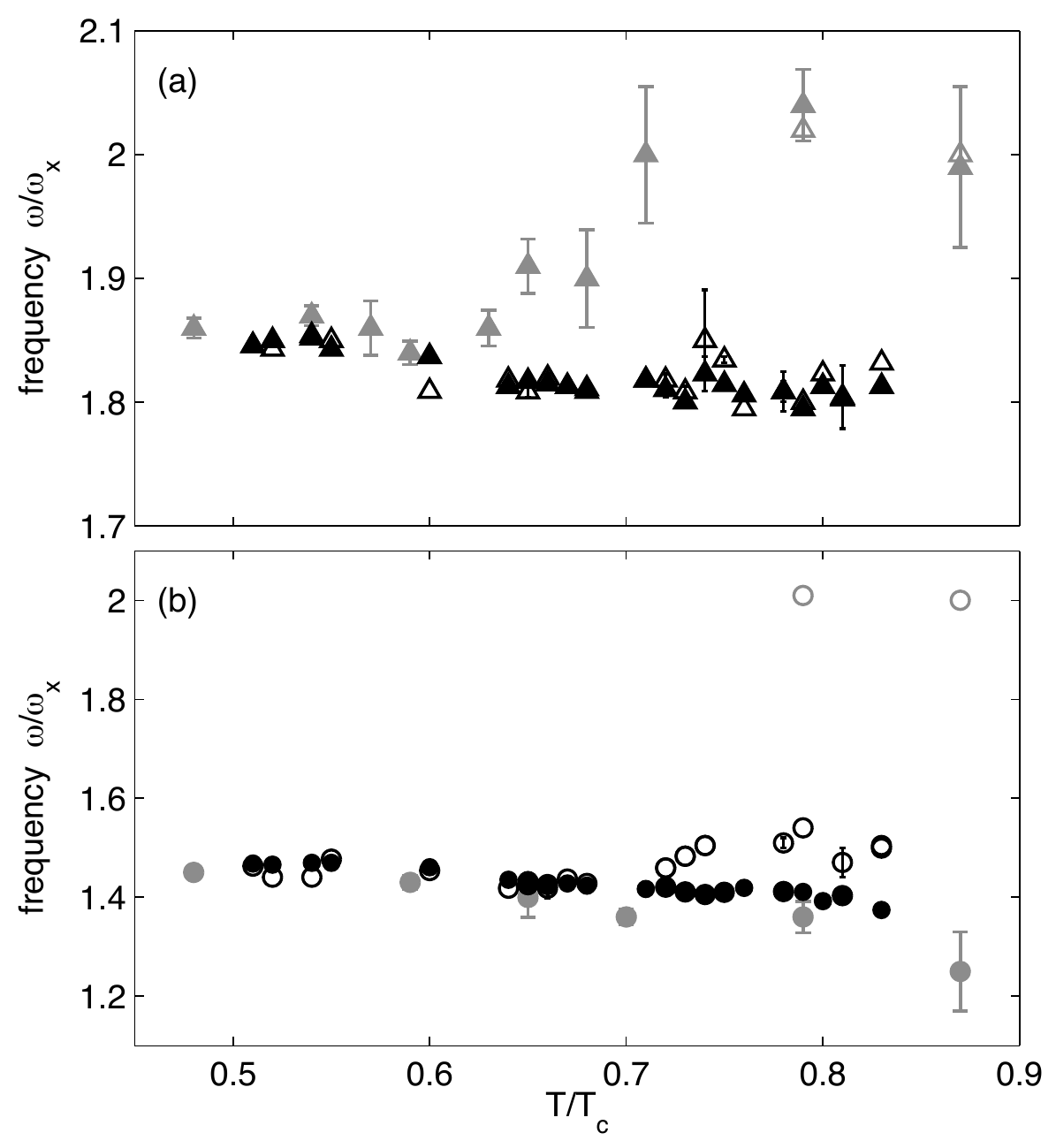}
\caption{\label{freq_results} Results for the frequency dependence on temperature for PGPE and experimental results of \cite{jin97}. (a) $m=0$ mode frequencies.  (b) $m=2$ mode frequencies. 
Experimental results (grey symbols) and PGPE results (black symbols). Frequency for condensate (solid symbols) and noncondensate (open symbols). Error bars on some PGPE results indicate the spread in values from different calculations at the same temperature.
}
\end{figure}

Our results for the mode frequency variation with temperature are presented in Fig. \ref{freq_results}, along with the experimental results from Jin \etal\ \cite{jin97} for comparison. 
We show results for the $m=0$ mode  and the $m=2$ mode, and give the frequencies for both the condensate (solid symbols) and noncondensate (open symbol) components.

\begin{figure}
\includegraphics[width=3.5in, keepaspectratio]{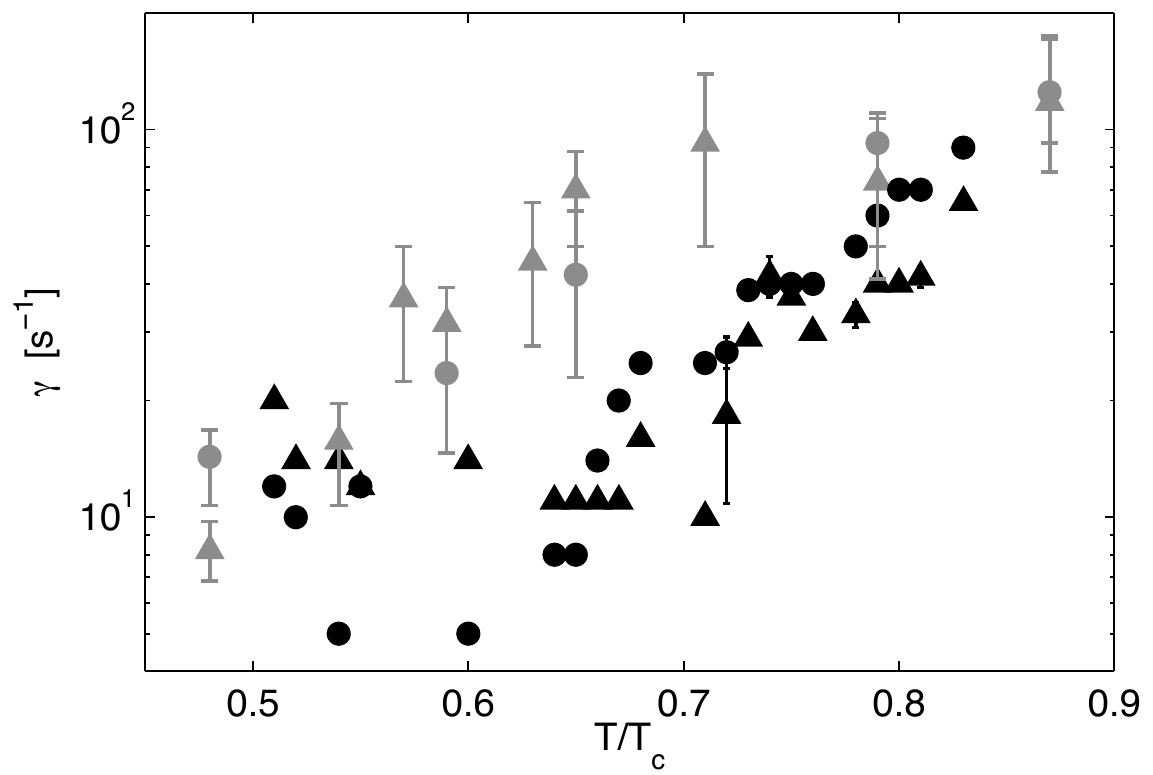}
\caption{\label{damping} Results for the damping rates variation with  temperature for PGPE and experimental results. Results are given for the condensate $m=0$ modes (triangles) and $m=2$ modes (circles). Experimental results (grey symbols) and PGPE results (black symbols).  Error bars on some PGPE results indicate the spread in values from different calculations at the same temperature.}
\end{figure}

We first examine the $m=0$ mode behaviour shown in  Fig. \ref{freq_results}(a).
At temperatures below $0.6 T_c$ the $m=0$ mode frequencies of the condensate and noncondensate components are almost the same, indicating that the two components oscillate together. In this temperature range the agreement with the experimental results for the condensate frequency is good. There are no experimental measurements for the noncondensate behavior in this regime as the noncondensate fraction is too small to measure. 
At temperatures above $0.6T_c$ our theoretical predictions and the experimental results exhibit markedly different behavior: As temperature increases above $0.6T_c$ our results (for both the condensate and noncondensate) decrease in frequency, whereas the experimental results show a rather rapid increase in frequency.  
This feature of the experimental results evaded theoretical description (e.g. see \cite{TheoryA3}) until the works of Jackson \etal\ \cite{TheoryB3}  in 2002 and Morgan \etal\ \cite{TheoryA5} in 2003.  We discuss the origin of the disagreement between PGPE and the experimental results further in Sec. \ref{sec:ecut_results}, and show that it arises from our lack of a dynamical description of the $\rI$ region. We note that the PGPE predictions of a downward trend in the frequency of the $m=0$ mode is consistent with the results of gapless Hartree-Fock-Bogoliubov calculations (see Fig. 2 of Hutchinson \etal\ \cite{TheoryA2}), indicating that anomalous average effects are included in our description. Our predictions are also in good agreement with the second order theory of  Morgan \etal\ for the mode frequency in the absence of direct thermal driving (see diamond symbols on Fig. 1(a) of Ref. \cite{TheoryA5}).

In Fig. \ref{freq_results}(b) the $m=2$ mode is considered. 
Here we see reasonable agreement between the PGPE predictions for the condensate oscillation frequency and the experimentally measured values at all temperatures simulated.  At high temperatures our predictions lie slightly above the experimentally measured values in a similar manner to the full second order predictions (i.e. including thermal driving) of Morgan \etal\ (see open circles in Fig. 1(b) of Ref. \cite{TheoryA5}).
For the $m=2$ mode noncondensate oscillation frequency, we see poor agreement with experimental results. There are no other theoretical predictions for the $m=2$ thermal modes for us to compare against as neither Ref. \cite{TheoryB3} or \cite{TheoryA5} present results for this case. The PGPE predictions for the noncondensate mode at temperatures above $0.70T_c$ show that this noncondensate decouples from the condensate, and that its frequency is well above that of the condensate. This behavior is qualitatively the same as that seen in experimental results (with experimental results only available at temperatures above $0.78T_c$), however the upward shift of the thermal mode frequency we calculate is much lower than that observed in experiments.
We discuss the origin of this quantitative disagreement between the PGPE and  experimental results further in Sec. \ref{sec:ecut_results}.

\subsubsection{Mode damping}
In Fig. \ref{damping} we present the PGPE predictions for the damping rates of the $m=0$ and $m=2$ condensate modes, which we compare against the experimental results. 
Although there is considerable scatter in the PGPE results, they appear to be consistent with the experimental measurements. In particular, we observe that in the temperature range $0.5T_c-0.6T_c$ the $m=0$ mode decays most rapidly (i.e. larger $\gamma$), while at higher temperatures the $m=2$ mode gradually takes over with a larger damping rate, broadly consistent with the experimental findings. 
We note that our choice of observable is more appropriate for determining mode frequency than decay due to the non-sinusoidal shape of the observable signal (see discussion in Sec. \ref{secobservables}). This will lead to a systematic shift in our predictions for the damping rate and may be responsible for the general downward shift of our results relative to the experimental measurements. In future work we will look into other observables to improve the accuracy with which we can analyse the mode damping rates.

\subsection{Cutoff dependence}\label{sec:ecut_results}
In this section we investigate the dependence of equilibrium and dynamic properties of the system on the energy cutoff ($\ecut$) used in our simulations. 

\subsubsection{Dependence of equilibrium states on $\ecut$}
To consider the effect of varying cutoff we follow the procedure discussed in Sec. \ref{sec:makeEq} to prepare an initial state with a cutoff of $\ecut=46\hbar\omega_x$, and equilibrium parameters of  $T=154$nK$=0.74T_c$,   $\Nc = 5.8\times10^3$, and $\nmin = 0.65$ where $\nmin$ is the mean occupation of the highest energy mode in the $\rC$ region (i.e. the least occupied $\rC$ region mode).  The quantity $\nmin$ is an important indicator of the PGPE validity, as it allows us to ensure that all the modes in $\rC$ are appreciably occupied.

To investigate the cutoff dependence we \emph{down-project} the equilibrium microstates $\cf$ of this system according to
\begin{equation}
\cfp = \mathcal{P}'\{\cf\},
\end{equation}
where $\mathcal{P}'$ is the projector for the cutoff $\ecut'<\ecut$.  The effect of this projection is to reduce the size of  $\rC$ to a smaller region, $\rC'$, and thus remove the occupation and energy of the modes lying between $\ecut'$ and $\ecut$. Since the constants of the motion, $E_{\rC'}$ ($<E_{\rC}$) and $N_{\rC'}$ ($<N_{\rC}$) have changed it is interesting to investigate if the equilibrium properties of the down-projected state differ from the original state. To check this we evolve $\cfp$ according to the PGPE  (\ref{PGPE}) (on region $\rC'$), and analyze the thermal state that $\cfp$ describes, after it is given time to thermalize. For the results we present here, we change $\ecut'$ from $45\hbar\omega_x$ to $30\hbar\omega_x$. Over this range the number of $\rC$ region modes decreases from
$5706$ (for the original state with $\ecut=46\hbar\omega_x$) down to $1575$ (for $\ecut'=30\hbar\omega_x$), i.e. the total number of $\rC$ regions modes changes by a factor of $3.6$ between the cutoff extremes we consider.

\begin{figure}
\includegraphics[width=3.5in, keepaspectratio]{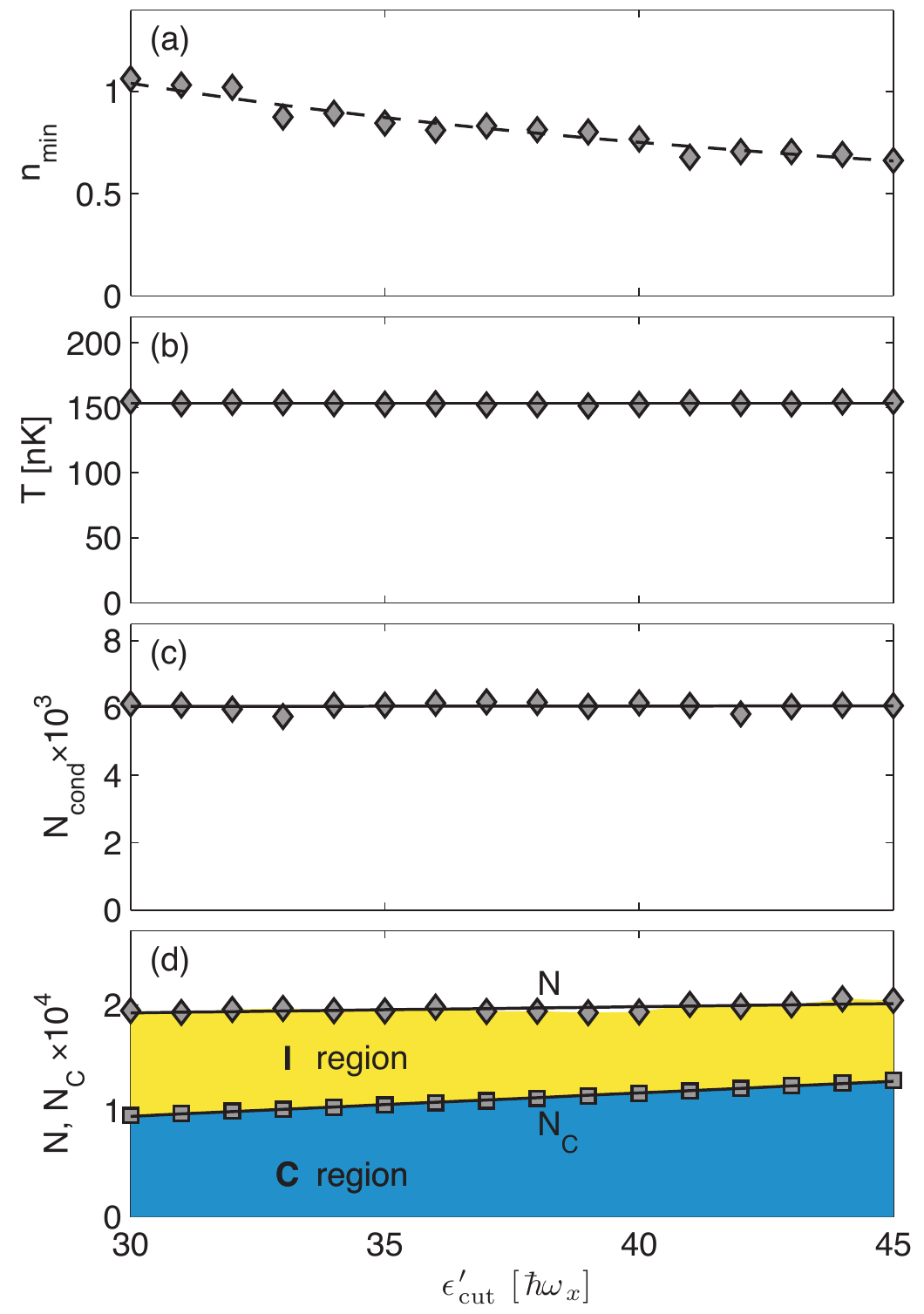}
\caption{\label{static_ecut} Dependence of equilibrium variables of down-projected state on the energy cutoff $\ecut'$. (a) Average occupation of highest energy $\rC$ region mode $\nmin$ (diamonds), (b) temperature $T$ (diamonds), (c) condensate number $\Nc$ (diamonds) and (d) total atom number $N$ (diamonds) and $\rC$ region atom number $N_{\rC}$ (squares).  Dashed line in (a) is a linear fit to $\nmin$ in the variable $1/\ecut'$, solid lines in (b)-(d) are linear fits to the data. In (d) the shaded regions indicate the relative number of atoms in the $\rC$ and $\rI$ regions for each value of $\ecut$. }
\end{figure}

In Fig. \ref{static_ecut} we present results for the equilibrium properties of our down-projected states. Fig. \ref{static_ecut}(a) shows the population ($\nmin$) of the highest harmonic oscillator state present in the simulation as a function of the energy cutoff. As $\ecut'$ is lowered, we see that the number occupying this highest state increases, in a manner consistent with the equipartition occupation of this mode  (i.e. $\nmin\sim k_BT/\ecut'$).

Figures \ref{static_ecut}(b)-(d) show the results for the macroscopic parameters $T$, $\Nc$, $N_{\rC}$, and $N$, respectively, of the down-projected state. We can see that these paramenters (excluding $N_{\rC}$) do not vary systematically with $\ecut'$, and conclude that the equilibrium parameters of our PGPE simulations are not dependent on the energy cutoff.  
These are the first results we are aware of showing the insensitivity of classical field method predictions to cutoff. Of course there are limits to how low we can take $\ecut'$, since our $\rC$ region must represent the condensate mode accurately which requires us to use a cutoff energy greater than the condensate chemical potential.

\subsubsection{Collective mode dependence on $\ecut$}\label{sseccollmodef}
Above we have shown that the equilibrium properties are insensitive to the cutoff defining the portion of the system in the $\rC$ and $\rI$ regions. In contrast we would expect that the PGPE theory for simulating collective modes, as developed in this paper, will show dependence on the cutoff. Fundamentally this is because the full dynamics of the $\rC$ region are simulated, while the  population of the $\rI$ region is neglected. Thus, in situations where the noncondensate dynamics are important 
the number of noncondensate modes included in $\rC$ will have a direct effect on the dynamical observables of the system. It would therefore seem desirable to include as much of the noncondensate population in the $\rC$ region as is possible, i.e. increase $\ecut$. However, there is a limit to how high we can set $\ecut$. As discussed in Sec. \ref{secPGPEformalism}, we formally require that all the $\rC$ modes are appreciably occupied for the classical field approximation to be a valid description of the Bose gas. For  $\ecut=46\hbar\omega_x$ we have $\nmin\approx0.65$, and so there is limited scope for using higher energy cutoffs.

In the absence of a dynamical theory for the $\rI$ region, adjusting the value of $\ecut$ allows us a mechanism by which to qualitatively investigate the role of the noncondensate dynamics in the collective mode dynamics. 
As we increase $\ecut$ from $30\hbar\omega_x$ to $46\hbar\omega_x$ the percentage of the total number of atoms in the $\rC$ region increases from approximately $50\%$ to $65\%$ (see Fig. \ref{static_ecut}(d)).

\begin{figure}
\includegraphics[width=3.5in, keepaspectratio]{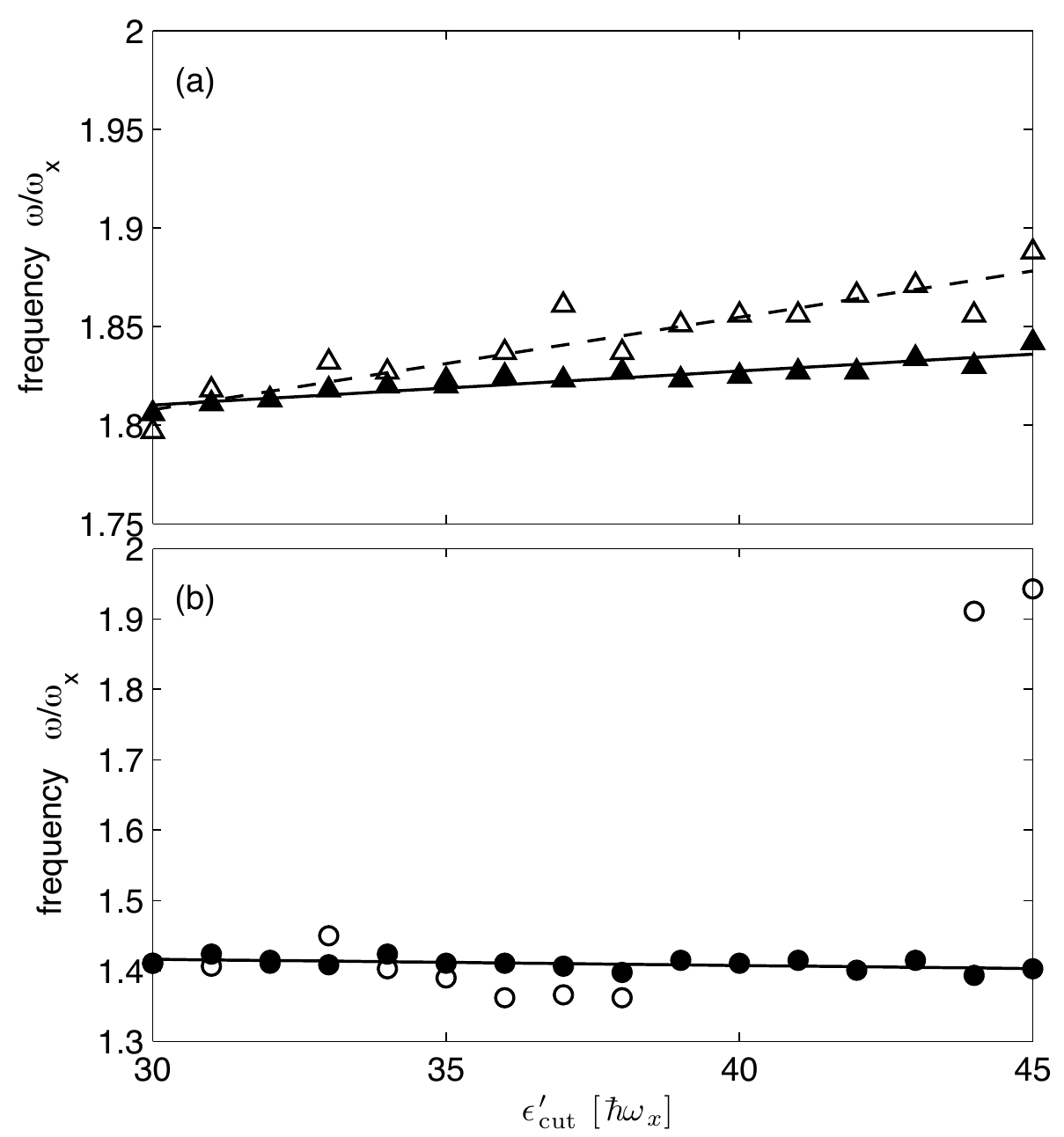}
\caption{\label{freq_ecut} Frequency dependence of mode excitation on energy cutoff, for  (a) $m=0$ mode and (b) $m=2$. Condensate frequency (solid symbols) and noncondensate frequency (open symbols). Solid lines are linear fits to the condensate data, and dashed line in (a) is fit to the noncondensate data. }
\end{figure}

We now investigate the frequency dependence of the $m=0$ and $m=2$ modes on the energy cutoff. Our procedure is the same as in Sec. \ref{secmodefreqs}, except that we consider a single temperature of $0.74T_c$ and sample our initial conditions for the PGPE from the equilibrium states 
of varying $\ecut'$ (i.e. those used to average for the macroscopic parameters shown in Fig. \ref{static_ecut}). From these simulations we determine frequencies of oscillation of the condensate and noncondensate components, with the results shown in Fig. \ref{freq_ecut}(a) and (b).  The frequency of the noncondensate $m=2$ mode for $\ecut'/\hbar\omega_x\in[38,44]$ are omitted as a single frequency fit of sufficient quality cannot be found (see discussion below).

The $m=0$ modes show a dependence on $\ecut'$, with the frequency of oscillation increasing as $\ecut'$ increases. The noncondensate frequency increases at a greater rate than the condensate, which is consistent with the increase in the condensate frequency arising from it being driven by the noncondensate component. Morgan has also seen this effect in his second order treatment by examining the influence of including thermal driving on the condensate mode (see Ref. \cite{Morgan_full}).  
Our results clearly indicate that including the dynamics of all noncondensate atoms is crucial to obtain a condensate mode frequency that would be comparable with the experimental results of Jin \etal\ \cite{jin97}. 

The $m=2$ condensate mode shows no almost dependence on the energy cutoff, suggesting that the noncondensate component does not couple strongly to this motion of the condensate. 
However, the noncondensate $m=2$ mode does show cutoff dependence:
At low $\ecut'$ (small thermal component) the noncondensate oscillates at the same frequency as the condensate, while at high $\ecut'$ (large thermal component) it oscillates at a frequency of $\omega\approx2\omega_x$ (which the expected value for the noninteracting limit of a thermal cloud). In the intermediate cutoff range, $\ecut'/\hbar\omega_x\in[38,44]$, a combination of the condensate dominated and noninteracting limit behaviors occur, and we were unable to fit a single frequency to these values. This suggests that there is a cutoff value, $\sim40\hbar\omega_x$, at which sufficient noncondensate is dynamically simulated for it to oscillate independently of the condensate.  

We now make some observations, from comparison of the goodness of the PGPE description of the experimental frequencies in Fig. \ref{freq_results} to the cutoff analysis of the modes in this section. For the $m=0$ modes, both the condensate and noncondensate predictions are in poor agreement with experiment at $0.74T_c$, and both are observed to be cutoff dependent [Fig. \ref{freq_ecut}(a)].  For the $m=2$ mode we find  that: (i) The condensate dynamics, which are in good agreement with experiment, are independent of cutoff [Fig. \ref{freq_ecut}(b)]. (ii) The noncondensate results, which are in poor agreement with the experiment at high temperatures, are strongly cutoff dependent [Fig. \ref{freq_ecut}(b)]. 
In general, these observations lead us to expect that cutoff independent predictions of the dynamical PGPE theory are likely to be accurate in the absence of a dynamical theory of the $\rI$ region, while cutoff dependent predictions are unreliable.  In the latter case a dynamical theory of the $\rI$ region is required.

\subsection{Dipole Mode}\label{sec:dipole_results}

\begin{figure}
\includegraphics[width=3.5in, keepaspectratio]{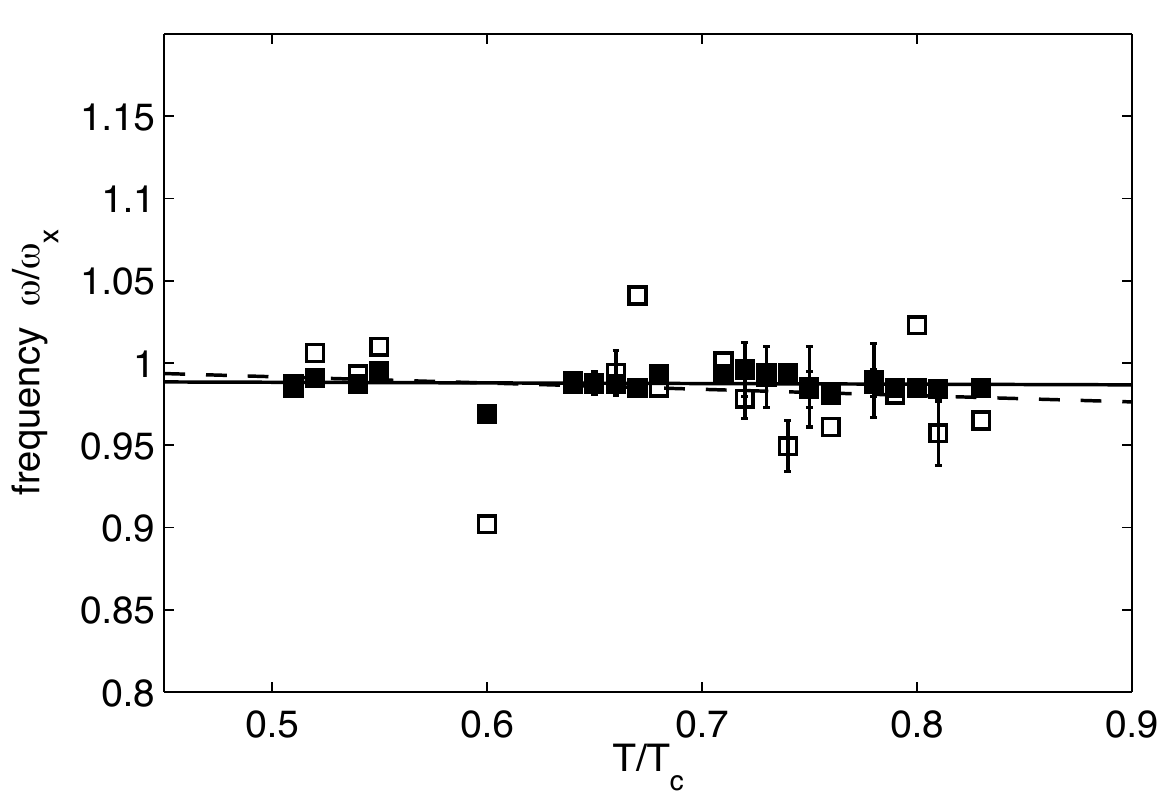}
\caption{\label{dipole} Frequency of dipole mode as a function of temperature. Condensate  (solid squares) and noncondensate (open squares). Error bars on some results indicate the spread in values from different calculations at the same temperature. Linear fit to the condensate (line) and noncondensate (dashed) results.}
\end{figure}

It is rigorously known that a harmonically trapped system will have a center-of-mass motion oscillation mode  at the trapping frequency (Kohn mode) \cite{KohnThrm}. This mode is an important test of theory and was analyzed in  experiment \cite{jin97} for the purposes of frequency calibration.  Due to the presence of a projector in the PGPE theory the Kohn mode is not a constant of motion (see Ref. \cite{Bradley}) and so for completeness we investigate the dynamics of this mode here.
 To do this we use the PGPE  (\ref{PGPE})  following the same procedure for setting up simulations as was done for the $m=0$ and $m=2$ modes, but with the dipole perturbation potential (\ref{pert2}). 
To analyse our data, we study the first moments $\langle p_x \rangle_{\sigma}$  for $\sigma=\{\rm{c},\rm{n}\}$ (see Sec. \ref{secobservables}) to provide observables for the condensate and noncondensate behavior. Our results, shown in Fig.\ref{dipole}, indicate that the condensate and noncondensate components both oscillate at approximately $1.0\omega_r$, as expected.  

\subsection{Relative phase of condensate-noncondensate oscillations}\label{sec:phase_results}
The relative phase of the condensate and noncondensate oscillations has played a central part in the explanation of the sharp jump in the frequency spectrum of the $m=0$ mode. Stoof and coworkers \cite{TheoryB1,TheoryB5} argued that the anomalous jump was caused by a transition from out-of-phase to in-phase oscillations of the condensate and noncondensate components at high temperature. Morgan \cite{Morgan_full} lends support to this theory by calculating the relative phase between the oscillations  of the condensate and noncondensate components, and shows that at moderate temperatures ($\sim0.5T_c$) the components oscillate out of phase, whereas at high temperatures ($\sim0.8T_c$) they oscillate in phase. This is consistent with the physical picture that a large noncondensate fraction oscillating at the noninteracting frequency $2\omega_x$ couples strongly to the condensate  $m=0$ mode and drives it at this higher frequency.  Morgan's results for the $m=2$ mode show the relative phase between the components increases with increasing temperature up to about $0.85T_c$, at which point a slight decrease is observed to begin. 

\begin{figure}
\includegraphics[width=3.5in, keepaspectratio]{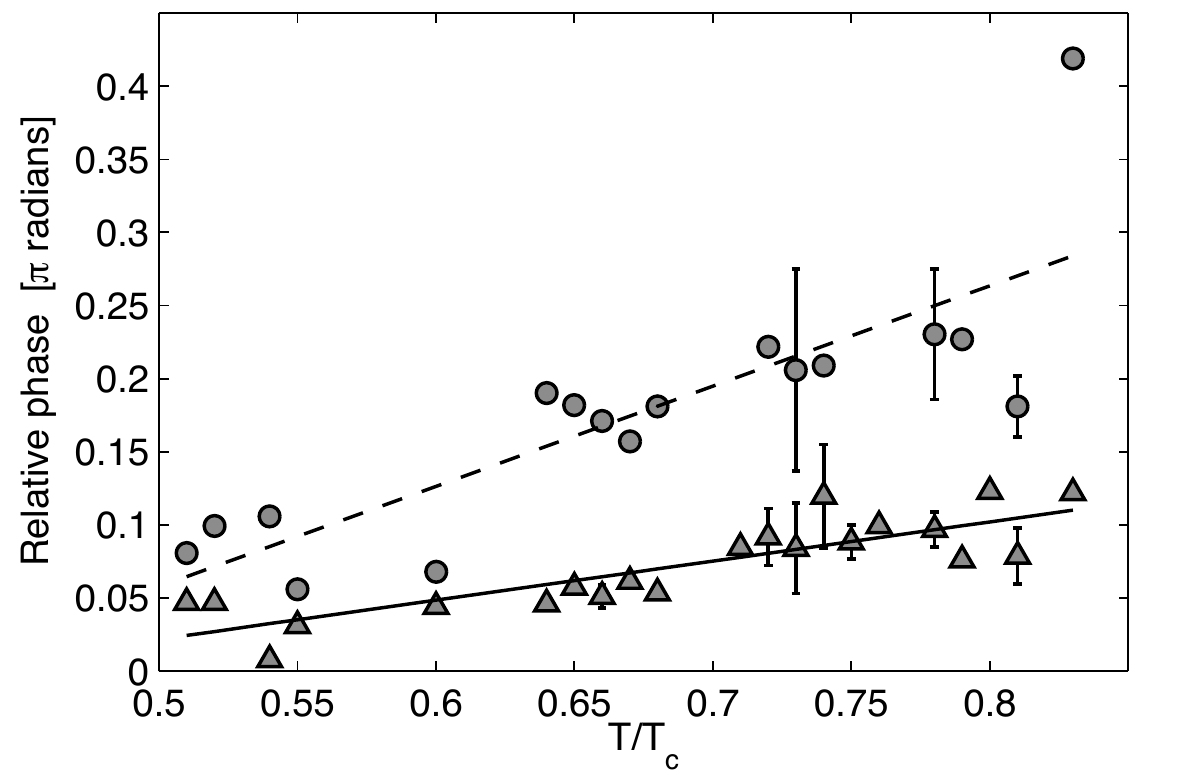}
\caption{\label{phase} Relative phase between the condensate and noncondensate  collective modes for the different symmetries studied:  $m=0$ mode  (triangles) and $m=2$ (circles). Error bars on some PGPE results indicate the spread in values from different calculations at the same temperature.   Solid line is a linear fit to the $m=0$ data and dashed line is a linear fit to the $m=2$ data. Same parameters as in Fig. \ref{freq_ecut}.}
\end{figure}

Here we follow the method of Morgan \cite{Morgan_full} closely. We calculate the phase difference between the two components using the first oscillation cycle after the perturbation is concluded. We find the relative phase by using the difference in minima of the two observable curves (see Fig. \ref{observablefigF}) as a fraction of the half period of the condensate oscillation, to give a result from zero to $\pi$. We present our results in Fig. \ref{phase}.  These  results are in reasonable qualitative agreement with those of Morgan at low temperatures, where the relative phases of each mode are increasing with temperature, with the $m=2$ mode having a larger phase angle to the $m=0$ mode at any given temperature (c.f. Fig. 11 \cite{Morgan_full}).  However, generally our predicted values for the relative phase are less than those calculated by Morgan, and more importantly, we do not see the sudden reduction in phase angle for the $m=0$ mode as temperatures increases above  $\sim0.7T_c$.
The likely explanation for our disagreement is that the fraction of noncondensate being dynamically simulated is not great enough so that: (i) At moderate temperatures $(\sim0.5T_c)$ 
the noncondensate is being dominated by the condensate oscillation, leading to a smaller than expected relative phase between the components (for both $m=0$ and $m=2$ modes). (ii) At higher temperatures $(>0.7T_c$) the noncondensate component is insufficiently dominant to effectively drive the condensate back in-phase with its natural oscillation (applying only to the resonantly coupled $m=0$ mode).  

\begin{figure}
\includegraphics[width=3.5in, keepaspectratio]{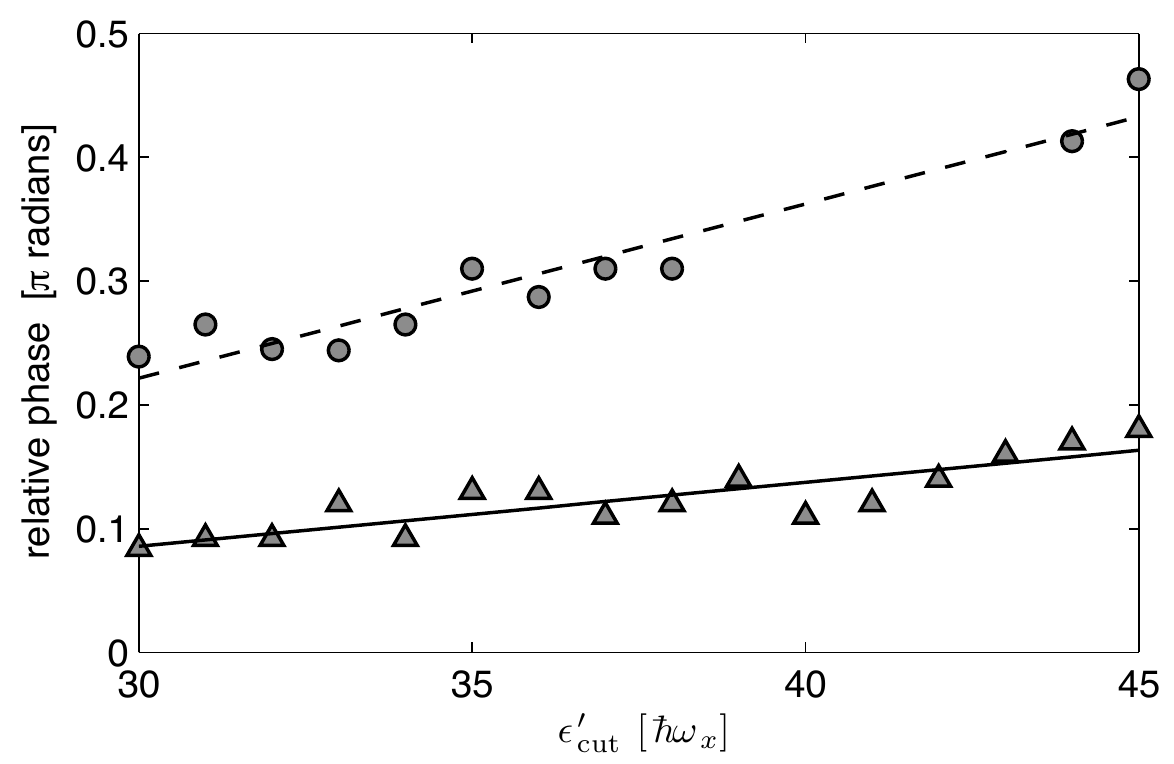}
\caption{\label{phase_ecut} Relative phase between the condensate and noncondensate collective modes the $m=0$ (triangles) and $m=2$ modes (circles) as a function of the down-projected energy cutoff $\ecut'$. Solid line is a linear fit to the $m=0$ data and dashed line is a linear fit to the $m=2$ data. Same parameters as in Fig. \ref{freq_ecut}.}
\end{figure}

 To further investigate these effects,  in Fig. \ref{phase_ecut} we show the dependence of the relative phase on the energy cutoff $\ecut'$, and hence noncondensate fraction in the $\rC$ region. In particular, we consider a system at temperature $0.74T_c$ for various $\ecut'$ (i.e. the same as was examined in Sec. \ref{sec:ecut_results}). At this temperature Morgan predicts a relative phase of $0.25\pi$ for the $m=0$ mode (and that with increasing temperature this phase decreases)  and a relative phase of $0.45\pi$ for the $m=2$ mode (which remains approximately constant with increasing temperature, before starting to decrease at about $0.85T_c$).
Our results for the dependence on $\ecut'$ shows that the relative phase of both modes increase with increasing cutoff, although the $m=0$ mode does so more slowly than the $m=2$ mode. The $m=2$ behavior indicates that as the noncondensate component being simulated increases (i.e.  as $\ecut'$ increases) its phase, relative to the condensate, becomes more independent.  We note that while the $m=2$ condensate mode is cutoff insensitive, the relative phase between the $m=2$ modes shows a dependence because the noncondensate mode does change character with $\ecut'$ (see Sec. \ref{sseccollmodef}).  The slower rate of increase in the the relative phase of the $m=0$ modes with cutoff (as compared to the $m=2$ modes, see Fig.  \ref{phase_ecut}) may be indicative of the resonant coupling between the components.

At our maximum value of cutoff ($\ecut=45\hbar\omega_x$), about $50\%$ of the noncondensate atoms are included in the PGPE description, and we speculate that a complete dynamical representation of the noncondensate would lead to this mode driving the condensate and the return to an in phase oscillation.

\section{Conclusions} \label{Conclusion}
We have presented a comprehensive study of the excitation spectrum of a Bose cloud at finite temperature by modelling the experiment of Jin \etal\ \cite{jin97} with the PGPE formalism. 
Our results for mode frequencies are in good agreement with experiment and other theories up to about $0.65T_c$. At temperatures above this our theory continues to provide a good description of the $m=2$ condensate mode. Currently our theory fails to predict the sudden increase in the frequency of the $m=0$ condensate mode at temperatures above $0.65T_c$. The origin of this failure in the current formalism is that we only provide a dynamical description for the portion of the noncondensate in the $\rC$ region. 

We have also examined the dependence of PGPE results on energy cutoff used to define the $\rC$ region. Importantly, we demonstrated the insensitivity of the equilibrium predictions to energy cutoff. 
The study of cutoff dependence in the collective mode results clearly reveals the importance of the interplay between condensate and noncondensate components in the mode behavior, and suggests a new practical validity check for the PGPE theory:  Dynamical predictions (in the absence of a dynamical theory for the $\rI$ region) should be verified to be independent of the cutoff energy.  
 
 The results of this first study with the PGPE give us great confidence that this theory is capable of providing a full description of the JILA experiments. To do this would require us to implement a dynamical description of the $\rI$ region, which we are currently pursuing.

\begin{acknowledgments} 
AB acknowledges support of a TEC Top Achiever Doctoral Grant.
PBB wishes to acknowledge useful discussions with A.~S.~Bradley, M.~J.~Davis, and D.~A.~W.~Hutchinson.
This work was supported by the New Zealand Foundation for Research, Science and
Technology under Contract Nos.  NERF-UOOX0703.  \end{acknowledgments}
 
\appendix
\begin{widetext}
\section{Parameters}\label{Aparams}
\begin{table}[htbp]
    \centering 
    \begin{tabular}{  |ccc|ccccc|  } % Column formatting, @{} suppresses leading/trailing space
     \hline
       $\quad\ecut \,\,[\hbar\omega_x]\quad$  & $\quad N_{\rC}\times10^3\quad$ &
       $\quad E_{\rC} \,\,[\hbar\omega_x]\quad$ & $\quad\Nc\times10^3\quad$ & $\quad N\times10^4\quad$ & $\quad T\,\, [$nK$]\quad$ & $\quad T/T_c\quad$ & $\quad{\nmin}\quad$  \\ 
       \hline     \hline
17 & 4.64 & 7.2 & 3.95 & 0.613 & 70 & 0.51 & 0.89  \\
17 & 4.54 & 7.25 & 3.74 & 0.64 & 78 & 0.55 & 1.12  \\
18 & 4.54 & 7.3 & 3.74 & 0.591 & 71 & 0.52 & 0.94  \\
19 & 4.54 & 7.5 & 3.67 & 0.598 & 74 & 0.54 & 0.89  \\
17 & 5.05 & 7.5 & 4.24 & 0.78 & 90 & 0.6 & 1.26  \\
32 & 11.9 & 12.3 & 7.46 & 1.75 & 127 & 0.64 & 0.89  \\
32 & 11.4 & 12.3 & 7.05 & 1.7 & 126 & 0.65 & 0.84  \\
32 & 11.1 & 12.3 & 6.8 & 1.7 & 126 & 0.65 & 0.79  \\
32 & 11.9 & 12.6 & 6.95 & 1.88 & 134 & 0.66 & 0.83  \\
32 & 10.9 & 12.3 & 6.26 & 1.62 & 126 & 0.66 & 0.81  \\
31 & 10.6 & 12.1 & 6.31 & 1.64 & 128 & 0.66 & 0.86  \\
30 & 10.6 & 11.9 & 6.52 & 1.75 & 133 & 0.67 & 0.91  \\
32 & 11.9 & 12.8 & 6.83 & 1.99 & 141 & 0.68 & 0.84  \\
36 & 11.9 & 14.4 & 5.6 & 1.95 & 143 & 0.71 & 0.81  \\
37 & 11.9 & 15 & 5.16 & 1.96 & 147 & 0.72 & 0.7  \\
36 & 11.9 & 14.6 & 5.64 & 2.01 & 149 & 0.72 & 0.85  \\
33 & 12.4 & 13.5 & 7.12 & 2.33 & 157 & 0.72 & 1.02  \\
35 & 12.4 & 14.4 & 6.22 & 2.27 & 158 & 0.73 & 0.91  \\
32 & 12.4 & 13.2 & 7.41 & 2.38 & 160 & 0.73 & 1.09  \\
46 & 13.1 & 22 & 5.77 & 2.03 & 154 & 0.74 & 0.67  \\
34 & 12.4 & 14.1 & 6.57 & 2.42 & 163 & 0.74 & 0.99  \\
38 & 11.9 & 16 & 4.92 & 2.15 & 159 & 0.75 & 0.88  \\
35 & 12.1 & 14.6 & 5.72 & 2.34 & 163 & 0.75 & 1.02  \\
33 & 12.4 & 13.8 & 6.89 & 2.51 & 168 & 0.76 & 1.01  \\
36 & 11.9 & 15.5 & 5.26 & 2.5 & 174 & 0.78 & 1.05  \\
33 & 12.4 & 14.1 & 5.98 & 2.79 & 179 & 0.78 & 1.15  \\
27 & 10.1 & 11.7 & 6.32 & 2.47 & 172 & 0.78 & 1.44  \\
32 & 12.4 & 13.8 & 7.06 & 2.98 & 186 & 0.79 & 1.29  \\
36 & 11.9 & 16 & 4.67 & 2.83 & 186 & 0.8 & 1.06  \\
37 & 11.9 & 16.5 & 4.62 & 2.72 & 185 & 0.81 & 1.03  \\
32 & 12.4 & 14.1 & 6.86 & 3.32 & 197 & 0.81 & 1.32  \\
25 & 9.59 & 11.3 & 6.32 & 2.96 & 190 & 0.81 & 1.79  \\
36 & 11.9 & 16.5 & 5.14 & 3.24 & 202 & 0.83 & 1.14  \\
      \hline\hline
    \end{tabular}
    \vspace*{3mm}
    \caption{A summary of the equilibrium state parameters used for the results reported in Fig. \ref{macroparams}  and used as initial states for the results presented in Sec. \ref{sec:freq_results}. The first three columns give the parameters used to generate the initial states and the remaining columns give the macroscopic parameters determined for these states.}
    \label{tab:PGPEtable}
 \end{table}
 \end{widetext} 
\bibliographystyle{apsrev}
\bibliography{excitations_paper}

\end{document}